\documentclass[aps,prc,onecolumn,superscriptaddress,showpacs,showkeys,nofootinbib]{revtex4-2}
\usepackage{graphicx}
\usepackage{bm}
\usepackage{epsfig}
\usepackage{amsfonts}
\usepackage{amssymb,amscd}
\usepackage{subfigure}
\usepackage{xcolor}
\definecolor{urlblue}{rgb}{0.2,0.4,0.7}
\usepackage{multirow}
\usepackage{comment}
\usepackage{amsmath}
\usepackage[linktocpage]{hyperref}
\hypersetup{
 linktocpage = false,
 urlcolor = urlblue,
 colorlinks = true,
 linkcolor = urlblue,
 anchorcolor = urlblue,
 citecolor = urlblue,
 pdfstartview = {XYZ null null 1.25} 
           }

\begin{document}

\title{Fully charmed tetraquark production in forward rapidity $pp$ collisions \\ at  LHC and FCC energies}

\author{Francesco G. {\sc Celiberto}}
\email{francesco.celiberto@uah.es}
\affiliation{Universidad
de Alcalá (UAH), Departamento de Física y Matemáticas, Campus Universitario, Alcalá de Henares, E-28805, Madrid, Spain}

\author{Andr\'e V. {\sc Giannini}}
\email{AndreGiannini@ufgd.edu.br}
\affiliation{
Federal University of Grande Dourados, 
Faculty of Exact Sciences and Technology,
Zip code 364, 79804-970, Dourados, MS, Brazil}
\affiliation{Departamento de F\'isica, Universidade do Estado de Santa Catarina, 89219-710 Joinville, SC, Brazil.}

\author{Victor P. {\sc Gon\c{c}alves}}
\email{barros@ufpel.edu.br}
\affiliation{Institute of Physics and Mathematics, Federal University of Pelotas, \\
  Postal Code 354,  96010-900, Pelotas, RS, Brazil}

\author{Yuri N. {\sc Lima}}
\email{ylima@if.usp.br}
\affiliation{Instituto de F\'isica, Universidade de S\~ao Paulo, C.P. 66318, 05315-970, S\~ao Paulo, SP, Brazil.}

\begin{abstract}
 In this paper, we investigate the   production of a fully charmed tetraquark state $T_{4c}$ in $pp$ collisions at forward rapidities through the fragmentation mechanism considering the Color Glass Condensate (CGC) formalism and the solution of the running coupling Balitsky-Kovchegov (BK) equation. The contributions of gluon -  and charm - initiated processes are taken into account, and the impact of an intrinsic charm component in the proton's wave function is estimated. Predictions for the   transverse momentum distribution of the $T_{4c}$ state are presented assuming different rapidities, distinct  quantum numbers of the state and center-of-mass energies. Our results indicate that the higher cross-section is associated with the production of a tensor state  $T_{4c}(2^{++})$, which is dominated by the gluon-initiated process. In contrast, the production of the axial-vector state $T_{4c}(1^{+-})$ is dominated by the charm-initiated process and is very sensitive to the presence (or not) of an intrinsic charm.
\end{abstract}

\maketitle
\date{\today}

\section{Introduction}
Since the proposition of the quark model in 1964, by Gell-Mann \cite{Gell-Mann:1964ewy} and Zweig \cite{Zweig:1964jf}, the possible existence of exotic hadrons, composed by four or five quarks, has been largely discussed in the literature (For reviews see, e.g., Refs.~\cite{Liu:2019zoy,Chen:2022asf}). In particular, exotic hadrons composed entirely of heavy quarks were originally investigated, more than four decades ago, in Refs.~\cite{Iwasaki:1976cn,Chao:1980dv,Ader:1981db}. Despite extensive experimental searches, the first candidate of a fully charmed tetraquark state, denoted $T_{4c}$, was only observed in 2020 by the LHCb Collaboration in the double $J/\psi$ invariant - mass spectrum \cite{LHCb:2020bwg}, and later confirmed by both the ATLAS  and CMS Collaborations \cite{ATLAS:2023bft,CMS:2023owd}. Such observations have motivated significant efforts focused on the description of the mass spectra and decay properties of $T_{4c}$, as well as on the understanding of its dynamical production mechanism, which still is an open question and remains a challenging problem in high energy physics. 

Over the last years, several approaches have been proposed to describe the $T_{4c}$ production in $pp$, $pA$, $AA$, $ep$, $e^+ e^-$ and $\gamma \gamma$ collisions  
\cite{Karliner:2016zzc,Berezhnoy:2011xy,Carvalho:2015nqf,Esposito:2018cwh,Bai:2016int,Wang:2020gmd,Maciula:2020wri,Feng:2020riv,  Zhang:2020hoh,Zhu:2020xni,Feng:2020qee,Goncalves:2021ytq,Biloshytskyi:2022dmo,Feng:2023agq,Abreu:2023wwg,Feng:2023ghc,Celiberto:2024mab,Bai:2024ezn,Belov:2024qyi,Bai:2024flh,Liang:2025wbt,Ma:2025ryo,Celiberto:2025dfe,Celiberto:2025ziy,Wang:2025hex}. In particular, the  nonrelativistic QCD framework (NRQCD)~\cite{Caswell:1985ui,Thacker:1990bm,Bodwin:1994jh} has been used to estimate the fully charmed tetraquark production in hadronic collisions, mainly motivated by the fact that its lowest Fock state, $|cc\bar{c}\bar{c}\rangle$, does not receive contributions from valence light quarks and dynamical gluons, justifying the factorization of long-distance dynamics. 
Such studies have been performed using the collinear factorization approach and have demonstrated that the fragmentation mechanism, where the $T_{4c}$ state is generated through the $g \rightarrow T_{4c}$ and $c \rightarrow T_{4c} $ transitions, becomes dominant for $p_T \gtrsim 20$ GeV{~\cite{Feng:2020riv,Celiberto:2025dfe,Celiberto:2025ziy,}, in analogy with heavy-quarkonium production~\cite{Braaten:1993rw,Cacciari:1994dr}}.
These analyses have focused on the $T_{4c}$ production at central rapidities, where the light cone momentum fractions of the colliding partons are similar ($x_1 \approx x_2 \approx M_{T_{4c}}/\sqrt{s}$). In contrast, at forward rapidities, the fully charmed tetraquark  production is expected to be dominated by collisions of projectile partons with large light cone momentum fractions ($x_1 \rightarrow 1$) with
target partons carrying a very small momentum fraction
($x_2 \ll 1$). In this kinematical region,  small-$x$ effects in the description of the QCD dynamics \cite{Gelis:2010nm}, and new large-$x$ contributions to the projectile wave function \cite{Brodsky:2015fna}, not taken into account in these previous studies, are expected to become important.

\begin{figure}[t]
\begin{tabular}{ccc}
\includegraphics[scale=0.27]{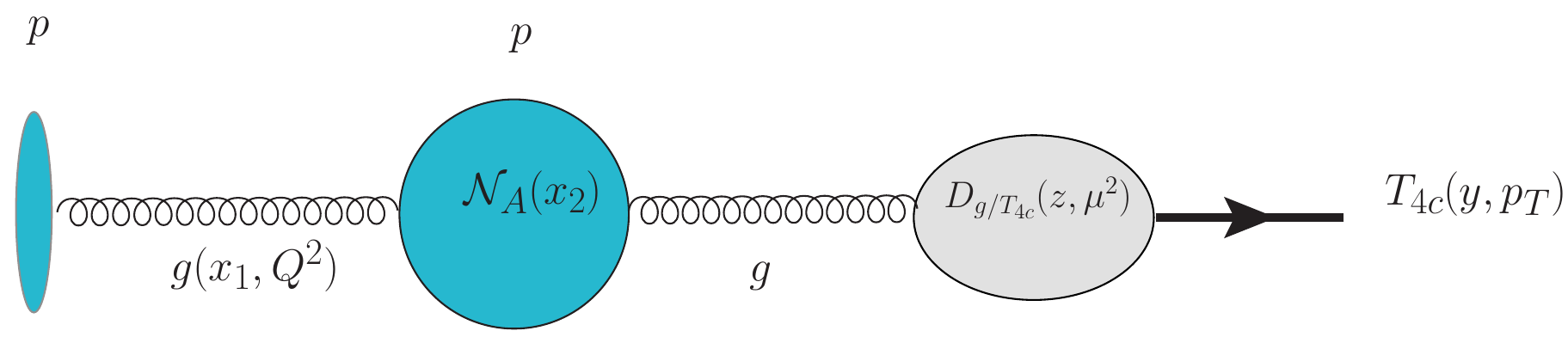} & \, & \includegraphics[scale=0.27]{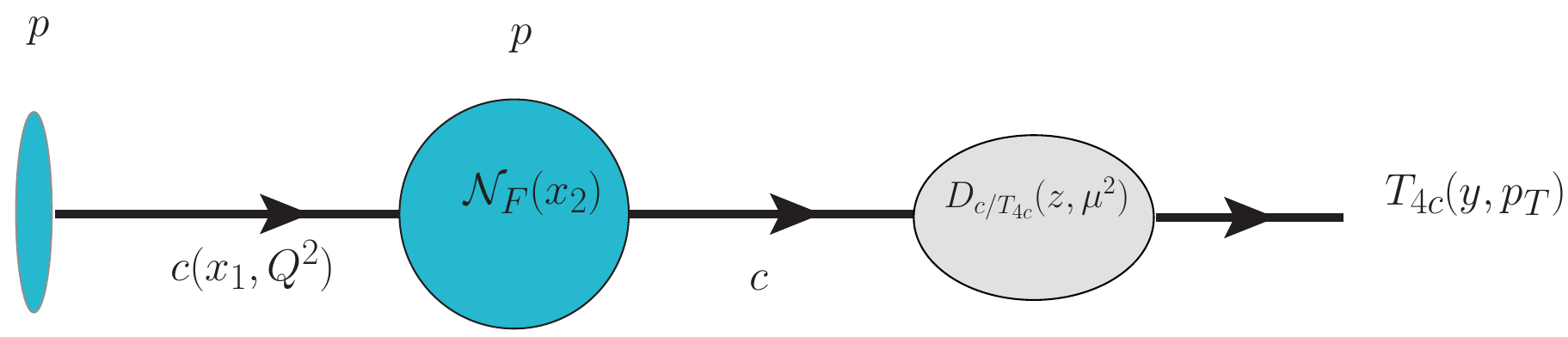}
\end{tabular}
\caption{Gluon (left)  and charm (right) - initiated contribution for the  production of a $T_{4c}$ state with transverse momentum $p_T$ and rapidity $y$ in $pp$ collisions.}
\label{Fig:diagram1}
\end{figure}

In this paper, we will investigate, for the first time, the impact of small and large-$x$ effects on the $T_{4c}$  production. In particular, we will consider the non-linear corrections to the QCD dynamics, as described by the Color  Glass Condensate (CGC) formalism \cite{CGC}, and the possibility that and intrinsic charm component could be present in the projectile wave function, which enhance the probability of finding a charm quark carrying a large momentum fraction $x_1$ of the projectile hadron. Our analysis is motivated by the recent evidences of an intrinsic charm (IC) component in the proton's wave function \cite{LHCb:2021stx,Ball:2022qks,NNPDF:2023tyk}
and by studies performed in Refs.~\cite{Goncalves:2008sw,Carvalho:2017zge,Giannini:2018utr,Maciula:2020dxv,Goncalves:2021yvw,Maciula:2022lzk} that  have demonstrated that this contribution can become dominant in the heavy meson production at ultra-forward rapidities. In particular, following Ref.~\cite{Lima:2024ksd}, we will estimate the contribution of the fragmentation mechanism for the  $T_{4c}$ meson production cross-section at forward rapidities using the CGC formalism and taking into account of gluon-initiated (GI) and charm-initiated (CI) contributions, which can be represented schematically as follows (See Fig. \ref{Fig:diagram1})
\begin{eqnarray}
\sigma(pp \rightarrow T_{4c} X) \propto  g(x_1,Q^2) \otimes {\cal{N}_A}(x_2) \otimes D_{g/T_{4c}} + c(x_1,Q^2) \otimes {\cal{N}_F}(x_2) \otimes D_{c/T_{4c}} \,\,,
\label{Eq:sig_imp}
\end{eqnarray}
where $g(x_1,Q^2)$ and $c(x_1,Q^2)$ are the collinear gluon and charm parton distribution functions (PDFs) of the projectile proton, $D_{g/T_{4c}}$ and $D_{c/T_{4c}}$ are  the collinear gluon and charm fragmentation functions (FFs) into the $T_{4c}$ state and the functions ${\cal{N}_A}(x_2)$ and ${\cal{N}_F}(x_2)$ are the adjunct and  fundamental forward scattering amplitudes, which encodes all the information about the hadronic scattering, and thus about the non-linear and quantum effects in the hadron wave function. Such amplitudes will be estimated using the CGC formalism, by solving the running coupling Balitsky-Kovchegov (rcBK) equation~\cite{BAL,KOVCHEGOV}. 
The first term in Eq.~\eqref{Eq:sig_imp} can be associated to the $gg \rightarrow g g$ subprocess and is expected to dominate at high center-of-mass energies and central rapidities. The charm initiated contribution is, in general, negligible in this same kinematical region. However, the recent results obtained in Ref.~\cite{Lima:2024ksd}   for the $D$-meson production at forward rapidities indicated that this contribution becomes important, especially if an intrinsic charm component is present. One of our goals is to investigate if this conclusion is also valid for the $T_{4c}$ case. In our analysis, we will consider that projectile PDFs are given by the  parametrizations derived by the CTEQ and NNPDF groups in Refs.~\cite{Hou:2019qau,Guzzi:2022rca,NNPDF:2021njg}, with and without the presence of an intrinsic charm component, and a detailed comparison between the associated predictions will be performed. Moreover, we will assume that the FFs for the transition of a charm or a gluon into a $T_{4c}$ state are described by the TQ4Q1.1 parameterization provided in Refs.~\cite{Celiberto:2024beg,Celiberto:2025dfe,Celiberto:2025ziy}, which allow us to derive predictions for the 
three lowest Fock-state $T_{4c}$ configurations: scalar ($J^{PC}=0^{++}$), axial-vector ($J^{PC}=1^{+-}$), and tensor ($J^{PC}=2^{++}$). Using these ingredients, we will derive predictions for the transverse momentum  distributions associated with the $T_{4c}$  production in $pp$ collisions at the energies of the Large Hadron Collider (LHC) and of the proposed Future Circular Collider (FCC)~\cite{FCC:2025lpp,FCC:2025uan,FCC:2025jtd}.

This paper is organized as follows. The next section presents a brief review of the formalism used to estimate the $T_{4c}$ meson production at forward rapidities together with the main ingredients in the calculations of the GI and CI contributions. 
In particular, we discuss the  distinct models for the treatment of the intrinsic charm component, the  initial condition of the BK equation used in our analysis, as well as the modeling of the FFs. 
In Section~\ref{Sec:Results} we present results for the transverse momentum distribution associated to the $T_{4c}$ meson production in $pp$ collisions at the LHC and FCC energies. 
Finally, in Section~\ref{Sec:conc}, we summarize our main results and conclusions.

\section{$T_{4c}$ production in the forward region at high energies}
\label{section:formalism}

In recent years, the $T_{4c}$ production in $pp$ collisions  has been investigated considering the collinear factorization, where all partons involved are assumed to be on mass shell, carrying only longitudinal momenta, and their transverse momenta are neglected in the QCD matrix elements \cite{Collins:1991ty}. However, the studies of heavy meson production in hadronic collisions, have demonstrated that at high energies the effects of a finite
transverse momenta of the incoming partons and  non-linear corrections to the QCD dynamics become important, motivating the proposition of generalized factorization schemes that  take into account of these effects (See, e.g., Refs.~\cite{Catani:1990eg,Tuchin:2004rb,Fujii:2005vj,Nikolaev:2005qs,Kovchegov:2006qn}). 
Following Refs.~\cite{Lima:2024ksd,Goncalves:2008sw}, we will consider the hybrid factorization scheme~\cite{Dumitru:2005gt,Altinoluk:2015vax}, which is adequate to describe  
the scattering of a dilute projectile off a dense target, which is the scenario present at forward rapidities.

As discussed in the introduction, we will  assume that the contribution associated with the fragmentation mechanism for the  $T_{4c}$  production will receive the contribution of GI and CI processes. 
For the GI contribution, represented in the left panel of Fig.~\ref{Fig:diagram1}, the differential cross-section for the production of a $T_{4c}$ state with transverse momentum $p_T$ at rapidity $y$ is expressed in the hybrid formalism by~\cite{Dumitru:2005gt,Altinoluk:2015vax}
\begin{eqnarray}
	\left.\frac{d\sigma_{pp \rightarrow T_{4c} X}}{dy d^2p_T}\right|_{GI} = \frac{\sigma_0}{2(2\pi)^2}\int_{x_F}^{1}dx_1\frac{x_1}{x_F}\bigg[g(x_1,Q^2)\,\tilde{{\mathcal{N}}}_A\left(\frac{x_1}{x_F}p_T,x_2\right)D_{g/T_{4c}}\left(z = \frac{x_F}{x_1},\mu^2\right)\bigg]\,\,,
\label{Eq:charm}	
\end{eqnarray}
where  $\sigma_0$ is a constant obtained by  fitting the HERA data (see below), $x_{1,2}$ represent the momentum fractions of the projectile and target partons that interact in the scattering process and $x_F$ is the Feynman-$x$ of the produced meson, which are defined by
\begin{equation}
	x_{1,2}=\frac{p_T}{z\sqrt{s}}e^{\pm y}
	\qquad\qquad \mbox{and} \qquad\qquad
	x_F = x_1 - x_2 \,\,.
\end{equation}
On the other hand, for the CI contribution, we will consider the approach proposed in Ref.~\cite{Goncalves:2008sw} and rederived in the hybrid formalism in 
Ref.~\cite{Altinoluk:2015vax}, which implies that the associated differential cross - section is given by~\cite{Goncalves:2008sw}
\begin{eqnarray}
	\left.\frac{d\sigma_{pp \rightarrow T_{4c} X}}{dy d^2p_T}\right|_{\rm CI} = \frac{\sigma_0}{2(2\pi)^2}\int_{x_F}^{1}dx_1\frac{x_1}{x_F}\bigg[c(x_1,Q^2)\,\tilde{{\mathcal{N}}}_F\left(\frac{x_1}{x_F}p_T,x_2\right)D_{c/T_{4c}}\left(z = \frac{x_F}{x_1},\mu^2\right)\bigg]\,\,.
\label{Eq:charm_CI}	
\end{eqnarray}
 The basic idea is that the incident parton (gluon or charm quark), present in the wave function of the incident proton, described by its PDFs, scatters off with the color background field of the target proton and then fragments into a $T_{4c}$ meson.
The interaction is described by the quantities  $\tilde{{\mathcal{N}}}_A$ and $\tilde{{\mathcal{N}}}_F$, which are the Fourier transform of the  adjunct and fundamental forward scattering amplitudes that satisfy the rcBK equation.  Finally, $D_{g/T_{4c}}$ and $D_{c/T_{4c}}$  are the gluon and charm FF in a $T_{4c}$ state. { Some comments are in order here. First, in this paper we are using the hybrid formalism at the leading order, which have been successful in describing, at least qualitatively, the different particle production spectra at forward rapidities. In recent years, different groups have worked on the calculation of the next - to - leading (NLO) corrections  to this formalism in order to achieve an accurate description of data. The first results have obtained uncontrolled large NLO corrections in the high transverse momentum region, where $p_T$ becomes much larger than the saturation scale, implying a negative cross - section~\cite{Chirilli:2011km,Chirilli:2012jd,Stasto:2013cha}. Such result has motivated more detailed studies~\cite{Altinoluk:2014eka,Watanabe:2015tja,Ducloue:2016shw,Iancu:2016vyg,Ducloue:2017dit,Xiao:2018zxf,Shi:2021hwx,Mantysaari:2023vfh}, which have proposed different approaches to solve this problem and that predict positive cross - sections at large - $p_T$, as the LO solution. In particular, the results derived in Ref.~\cite{Shi:2021hwx}  using the resummation formalism indicate that the inclusion of these NLO corrections increases the normalization of the predictions for the light hadron production at forward rapidities in comparison to the LO results, improving the description of data. However, the impact of the NLO corrections on the heavy hadron production is still an open question that deserves a more detailed analysis. A second comment is that our focus will be on $pp$ collisions, but the formalism presented in this paper can be generalized for a nuclear target (e.g. $pPb$ collisions) by considering the solution of the BK equation for the nuclear case. However, in this case, the results derived e.g. in Ref.~\cite{Mantysaari:2023vfh} indicate that in order to describe the nuclear modification factor $R_{pPb}$, and obtain the expected behavior at large $p_T$, $R_{pPb} \rightarrow 1$, the NLO corrections must be taken into account. Finally, another important aspect that must be emphasized is that some further theoretical efforts~\cite{Altinoluk:2014oxa,Altinoluk:2015gia,Chirilli:2018kkw,Altinoluk:2020oyd,Altinoluk:2024dba,Altinoluk:2024tyx} have been recently performed to go beyond the eikonal approximation assumed in the LO and NLO calculations discussed above.  These studies indicated that such next - to - eikonal corrections are important to match the hybrid formalism with the collinear approach at large - $p_T$. In principle, the inclusion of these corrections are expected to modify our predictions, but its impact is still unclear. Surely, such topic deserves to be investigated in the future. In what follows, we will describe in more detail the ingredients used in our calculations, which are performed at leading order and that assume the validity of the eikonal approximation.}

\vspace{0.25cm}
{\bf The parton distribution functions.} Over the last decade, the combination of the  high precision measurements performed in $ep$ collisions at HERA with the data obtained at the LHC has allowed largely improve the description of the proton structure~\cite{Gao:2017yyd}. Such results imply that the gluon distribution dominates at small values of the Bjorken $x$ variable, while for large $x$ the proton structure is determined by valence quarks. In addition, sea quarks created by the QCD evolution are also present. In particular, such evolution also generates $c\bar{c}$ pairs, implying a charm content within the proton, usually denoted  extrinsic charm component. Recent results also indicate the presence of an intrinsic charm component in the proton wave function~\cite{LHCb:2021stx,Ball:2022qks,NNPDF:2023tyk}. 
The existence of the intrinsic charm (IC) component was first proposed long
ago by Brodsky, Hoyer, Peterson and  Sakai (BHPS) in Ref.~\cite{Brodsky:1980pb} and since then other models for IC have been discussed \cite{Navarra:1995rq,Paiva:1996dd,Hobbs:2013bia,Pumplin:2007wg}.
The basic idea is that the proton light cone wave function has higher Fock states, one of them being $|qqqc\bar{c}\rangle$, with the probability of finding this configuration given by the inverse of the squared invariant mass of the system. Moreover, the presence of an IC component implies a large enhancement of the charm distribution at large $x$ $(> 0.1)$, which has impact on the production of heavy mesons at forward rapidities. 
The studies performed in Refs.~\cite{Lima:2024ksd,Goncalves:2008sw,Carvalho:2017zge,Giannini:2018utr,Maciula:2020dxv,Goncalves:2021yvw,Maciula:2022lzk}, for the case of $D$ meson production in $pp$ collisions, have demonstrated that the charm initiated a contribution is negligible for central rapidities, but becomes important for forward rapidities and can dominate if an intrinsic component of charm is present in the proton's wave function. 
Such results motivate the analysis of the impact of the IC component on the $T_{4c}$ production.

\begin{figure}[t]
\includegraphics[scale=0.4]{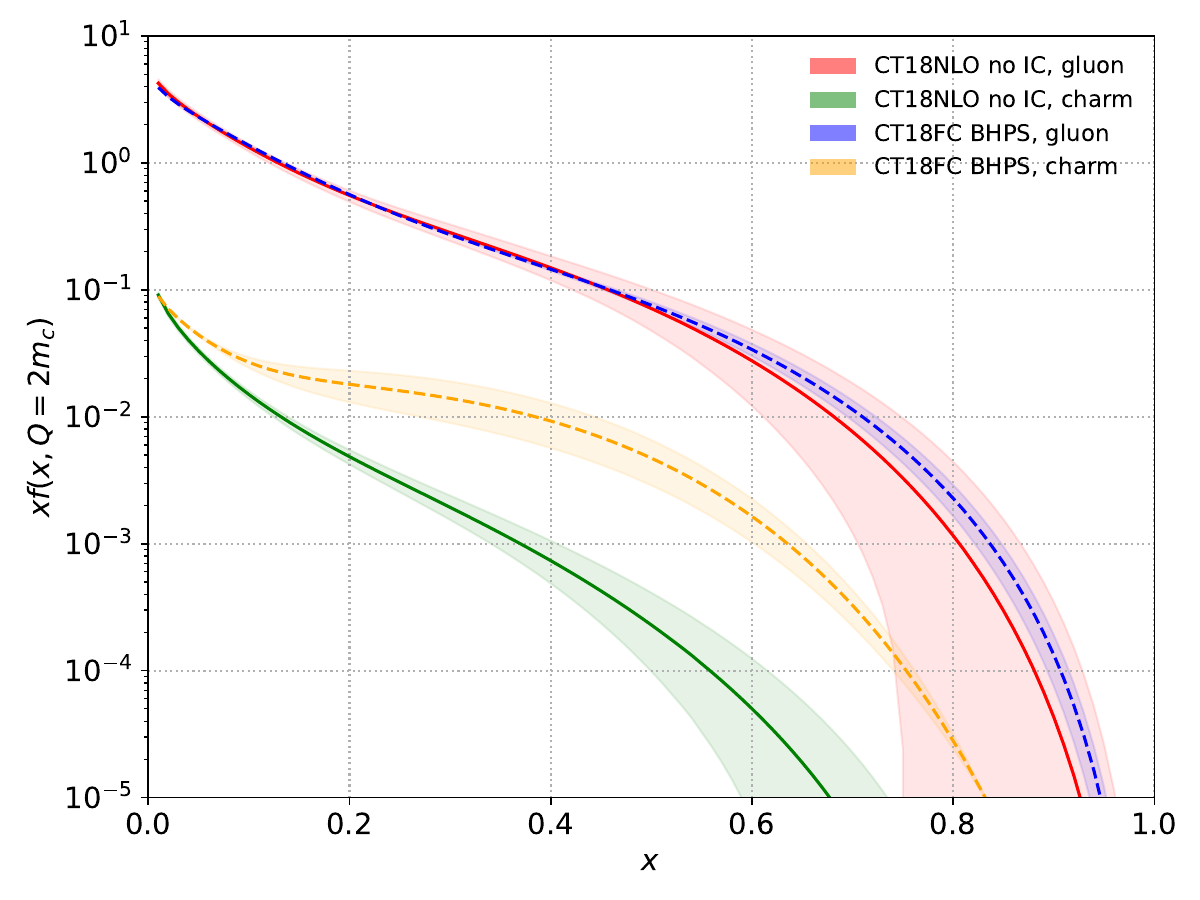}
\includegraphics[scale=0.4]{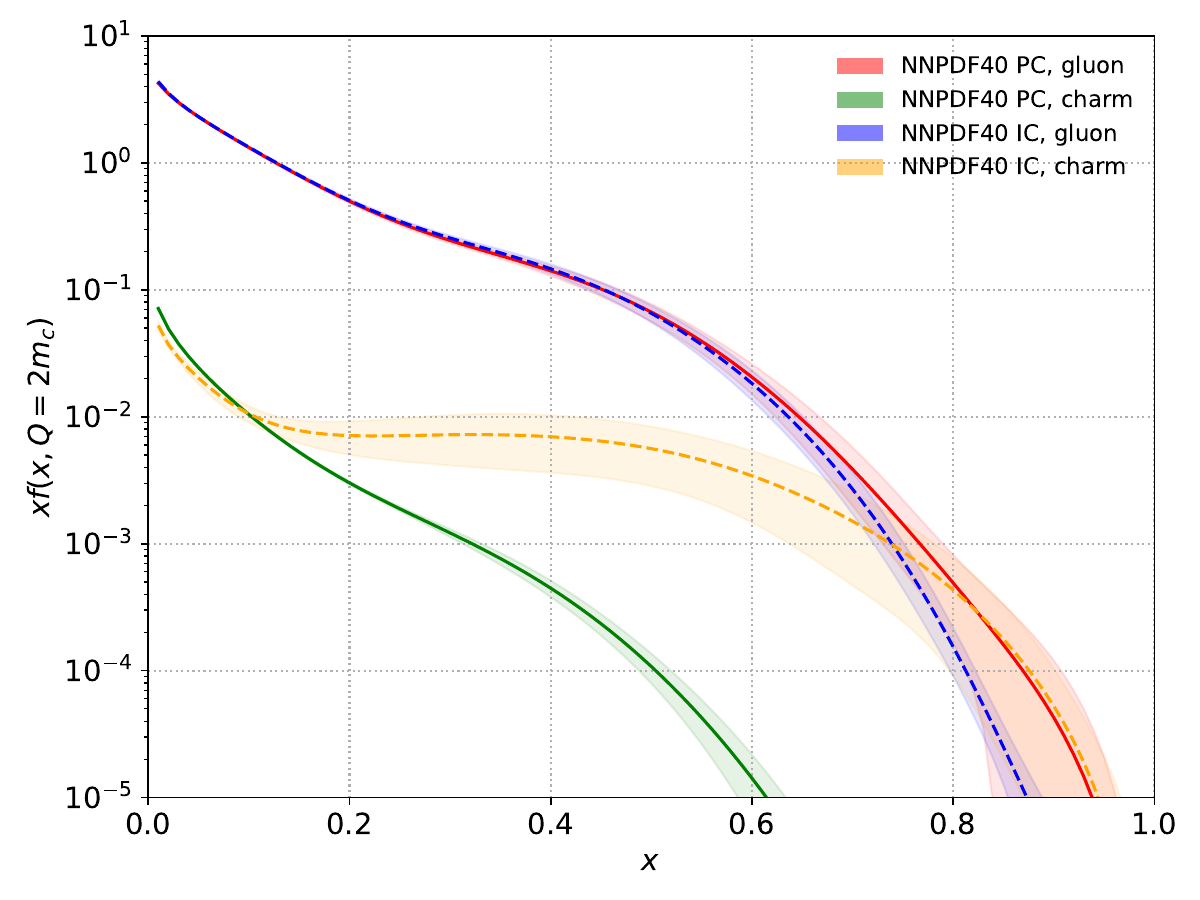}
\caption{Predictions of the different intrinsic charm models for the $x$ - dependence of the charm (lower blue curves) and gluon (upper red curves) distributions, as obtained by the CT18 (left panel) and NNPDF (right panel) parametrizations~\cite{Hou:2019qau,Guzzi:2022rca}. }
\label{Fig:pdfs}
\end{figure}

 In our analysis, we will consider the PDFs obtained in the global analysis performed in  
 Refs.~\cite{Hou:2019qau,Guzzi:2022rca,NNPDF:2021njg}, with and without the presence of an intrinsic charm component in the proton wave function. Although these two groups have basically used  the same set of data, the  parameterizations have been derived using distinct methodologies and 
 different assumptions for the initial conditions. As a consequence, they predict different behaviors for the charm and gluon distributions, as can be verified in Fig.~\ref{Fig:pdfs}, where we present a comparison between the charm and gluon PDFs predicted by the CT18 (left panel) and NNPDF (right panel) parameterizations for a fixed hard scale $Q = 2m_c$. 
 The parameterizations that assume a presence of an IC component (CT18FC BHPS and NNPDF40 IC) imply an  enhancement of the charm distribution at large $x$ $(> 0.1)$ (lower curves in Fig.~\ref{Fig:pdfs}) in comparison to the no IC predictions (CT18NLO no IC and NNPDF40 PC). We have that  the corresponding gluon distributions are also modified by the inclusion of intrinsic charm, as observed in the upper curves of Fig.~\ref{Fig:pdfs}. Moreover, we have that the amount of charm quarks at large $x$ is larger in the NNPDF40 parameterization in comparison to the CT18FC one. In the next section, we will calculate the $T_{4c}$ production at large rapidities assuming these distinct parameterizations, which will allow us to estimate the dependence of our predictions on the choice of PDF for the projectile proton, as well as the impact of an intrinsic charm component on the transverse momentum distributions.
 {The bands in Fig.~\ref{Fig:pdfs} illustrate the PDF-related uncertainties at the input energy scale $Q = 2m_c$.}

\begin{figure}[t]
\begin{tabular}{ccc}
\includegraphics[scale=0.37]{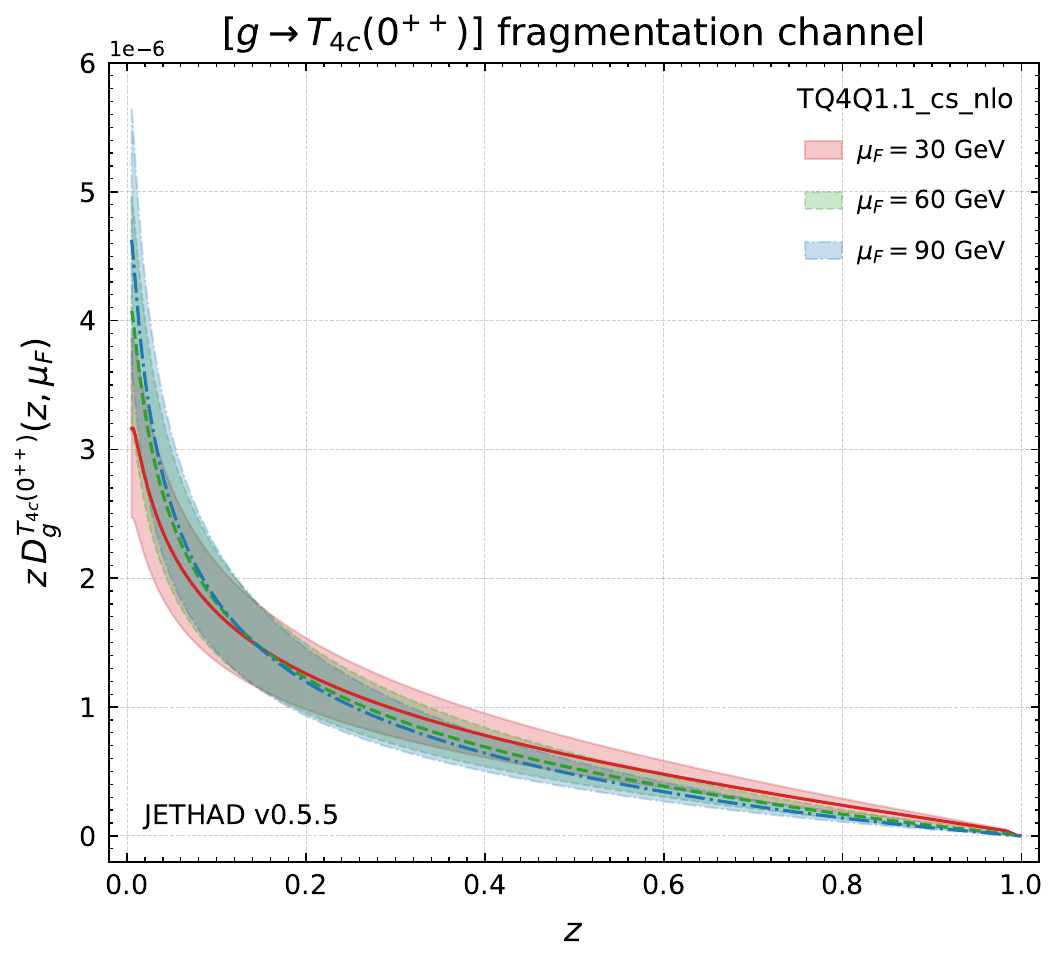} & \;\; & \includegraphics[scale=0.37]{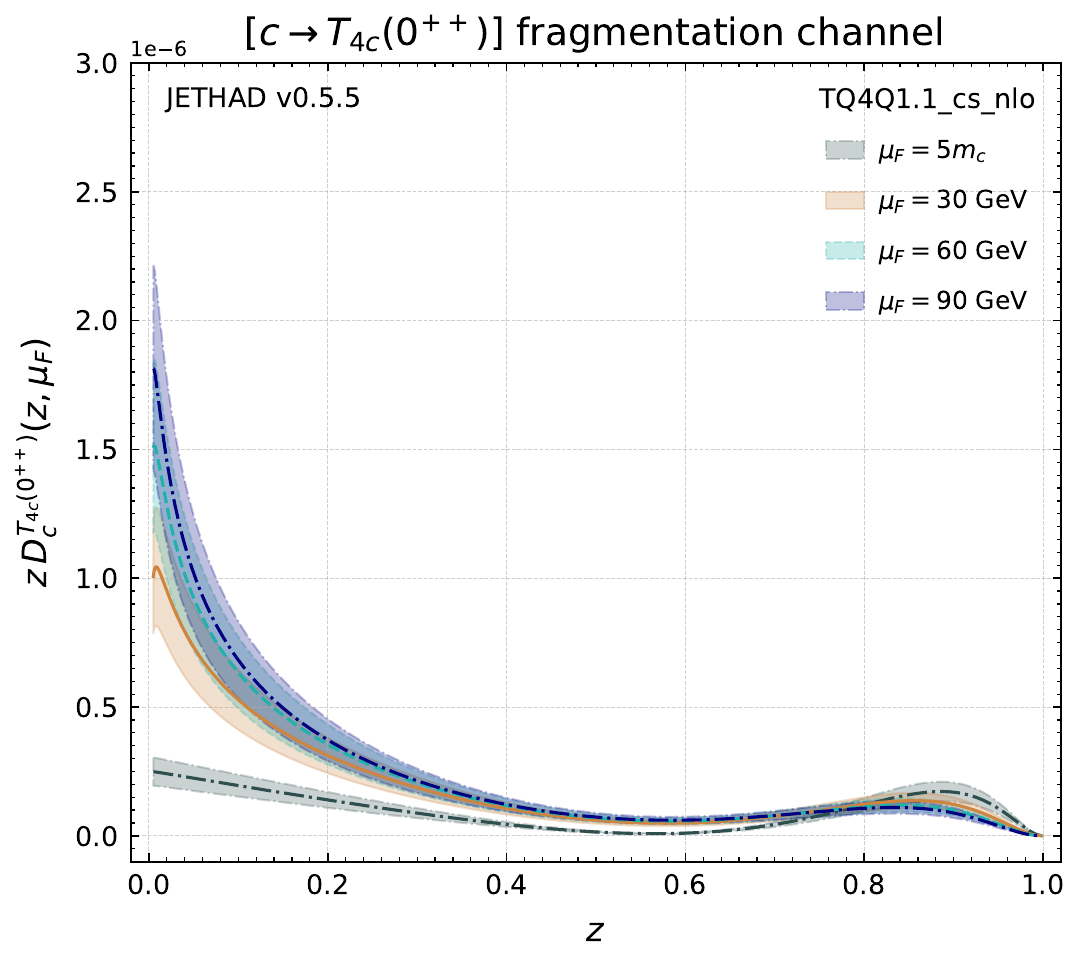} \\
\includegraphics[scale=0.37]{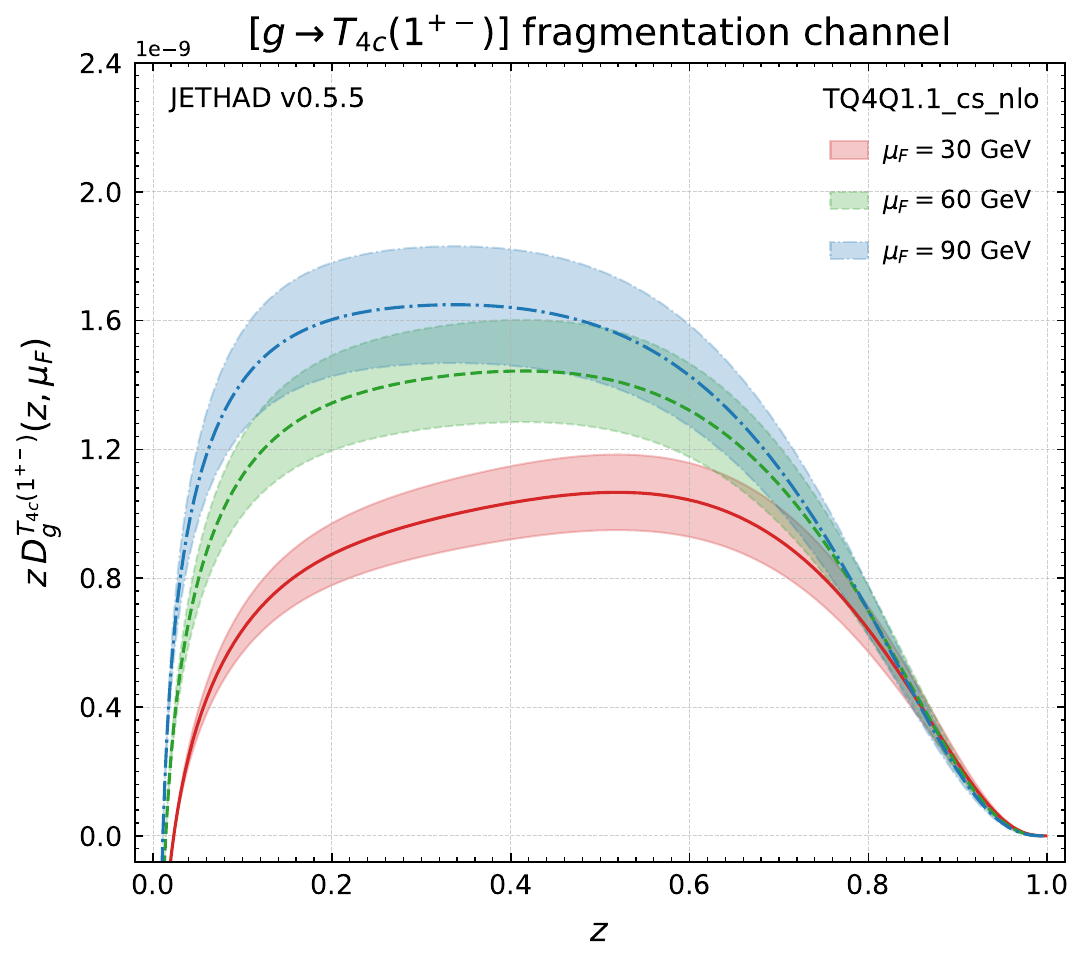} & \;\; & \includegraphics[scale=0.37]{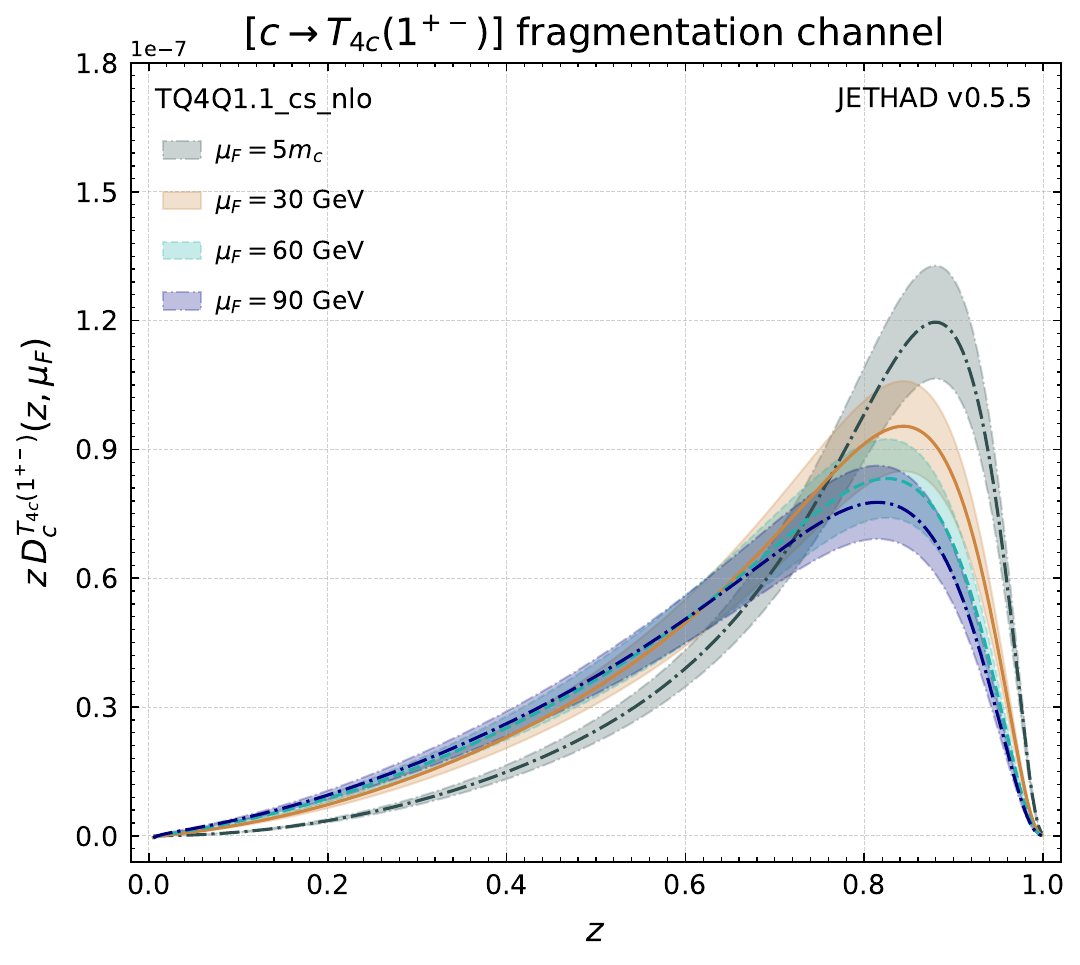}\\
\includegraphics[scale=0.37]{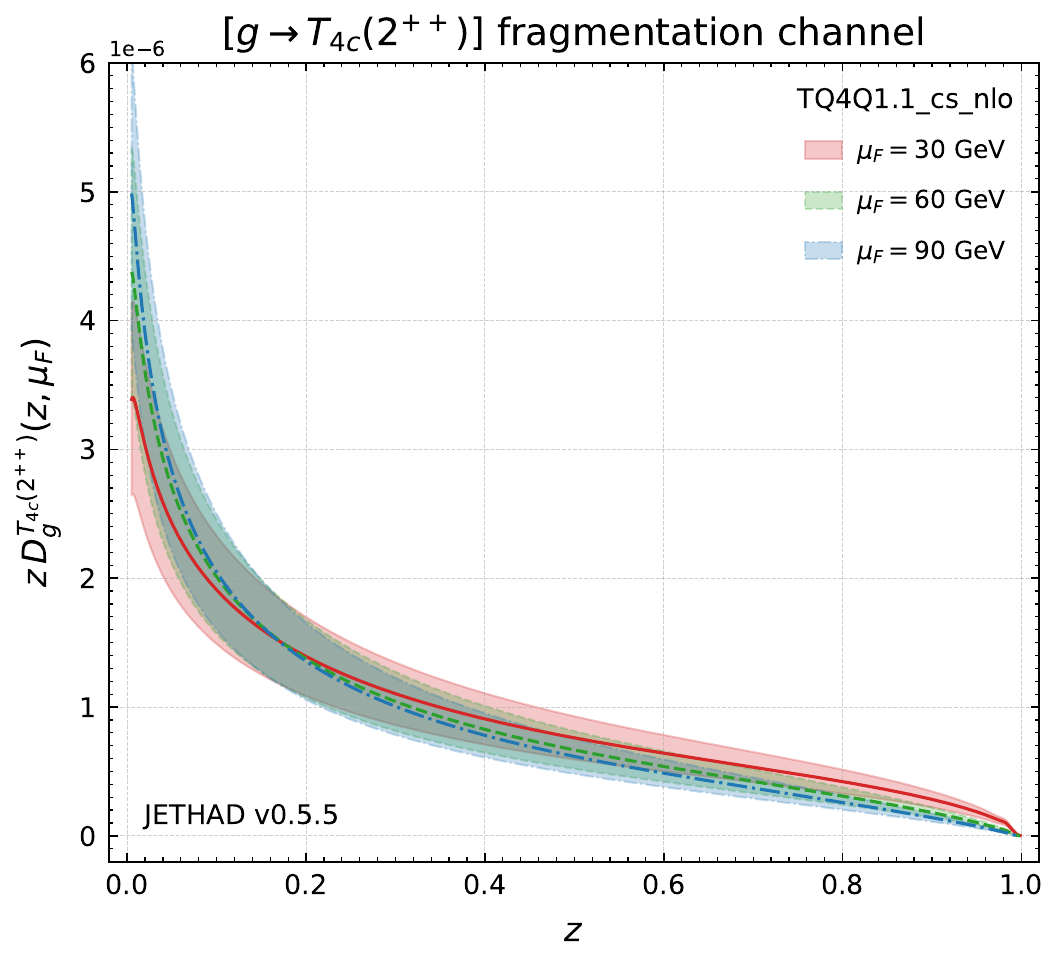} & \;\; & \includegraphics[scale=0.37]{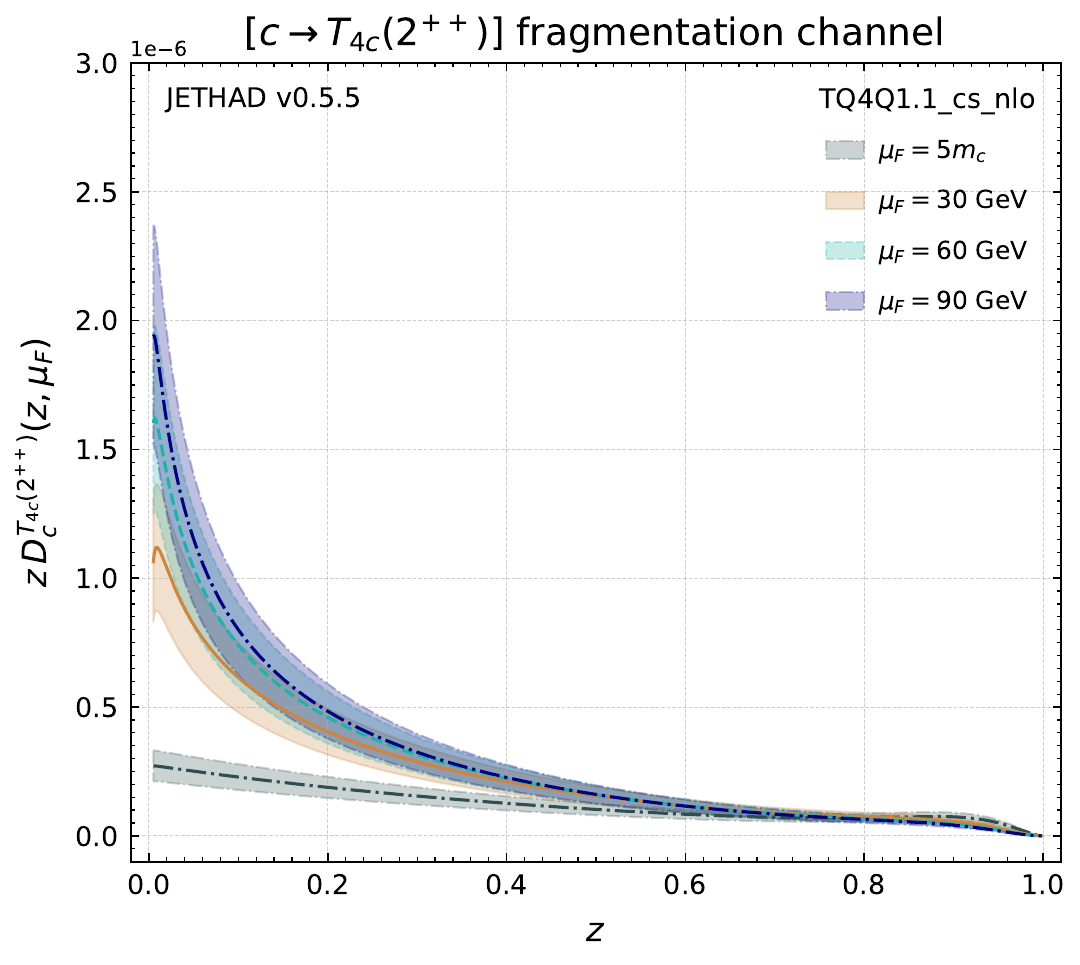}
\end{tabular}
\caption{Dependence on $z$ of the TQ4Q1.1 functions describing the gluon (left panels) and charm (right panels) fragmentation channels into scalar tetraquarks $T_{4c}(0^{++})$ (upper panels), axial-vector tetraquarks $T_{4c}(1^{+-})$ (middle panels) and tensor tetraquarks $T_{4c}(2^{++})$ (lower panels), calculated at various factorization scales $\mu_F$.
Filled bands represent the uncertainties associated with LDMEs.}
\label{Fig:frags}
\end{figure}
\vspace{0.25cm}
{\bf The scattering amplitudes.}
 For the adjunct and fundamental forward dipole scattering amplitude,  we will consider that both quantities are related in the impact parameter space by the relation ${{\mathcal{N}}}_A (x,r) = 2 {{\mathcal{N}}}_F (x,r) - {{\mathcal{N}}}_F^2 (x,r) $, where $r$ is the dipole size, and estimate ${{\mathcal{N}}}_F (x,r)$  by solving the  BK equation~\cite{BAL,KOVCHEGOV} including running coupling corrections~\cite{Albacete:2007yr}, denoted rcBK, which represents the simplest non-linear evolution equation for the dipole-hadron scattering amplitude, being actually a mean field version of the first equation of the Balitsky hierarchy~\cite{BAL}. We will consider the solution obtained assuming an initial condition inspired by the MV~\cite{McLerran:1997fk} model, given by  
\begin{gather}
{\mathcal{N}}_F(x_0 = 0.01,r) = 1 - \exp\bigg[-\frac{(r^2 Q_{s,0}^2)^{\gamma}}{4} \ln \left(\frac{1}{\Lambda r}+ e\right)\bigg] \qquad \text{(MV-like)}\,,
\end{gather}
where $\Lambda = \Lambda_{QCD}(n_f =3)$  and the free parameters, the initial saturation scale, $Q_{s,0}$ and the anomalous dimension, $\gamma$, have been determined by fitting the HERA data~\cite{Albacete:2009fh,Albacete:2010sy}. 
As in Refs.~\cite{Lima:2024ksd,Lima:2022mol,Lima:2023dqw}, we will assume a MV-like initial condition with $Q_{s,0}^2 = 0.1597$ GeV$^2$ and $\gamma = 1.118$. Moreover,  we will assume the values of $\sigma_0$ obtained in Ref.~\cite{Albacete:2010sy}. For a more detailed discussion about the scattering amplitudes, we refer the interested reader to the Ref.~\cite{Lima:2024ksd}.

\vspace{0.25cm}
{\bf The TQ4Q1.1 fragmentation functions.}  
Finally, the gluon and charm FFs employed in this work are taken from the recent TQ4Q1.1 determinations~\cite{Celiberto:2024beg,Celiberto:2025dfe,Celiberto:2025ziy}.  
These functions describe the collinear fragmentation of gluons and heavy quarks into fully heavy tetraquarks in the three lowest Fock-state configurations: scalar ($J^{PC}=0^{++}$), axial-vector ($J^{PC}=1^{+-}$), and tensor ($J^{PC}=2^{++}$).  
They are constructed from initial-scale inputs~\cite{Feng:2020riv,Bai:2024ezn} calculated within the Non-Relativistic QCD (NRQCD) effective theory~\cite{Caswell:1985ui,Thacker:1990bm,Bodwin:1994jh}, assuming single-parton fragmentation at leading power, and evolved through a threshold-consistent next-to-leading-order DGLAP evolution implemented in the Heavy-Flavor Non-Relativistic evolution (HF-NRevo) scheme~\cite{Celiberto:2025euy,Celiberto:2024mex}.  
This hybrid evolution framework ensures smooth matching across heavy-flavor thresholds and a consistent treatment of multi-flavor dynamics.  
Moreover, the TQ4Q1.1 release provides a systematic assessment of theoretical uncertainties.

These properties make the TQ4Q1.1 sets particularly well suited for applications in the forward-rapidity regime, where hybrid (CGC-based) factorization applies.  
In this kinematic region, the hard partonic subprocess and the subsequent collinear fragmentation are cleanly factorized: the small-$x$ evolution of the dense target is encoded in the unintegrated gluon distribution, while the moderate-$x$ projectile dynamics is naturally described by the collinear FFs.  
This complementarity allows the TQ4Q1.1 determinations to provide reliable nonperturbative inputs for the description of forward exotic-hadron production within the hybrid formalism.
The TQ4Q1.1 sets are part of a broader program dedicated to the high-energy study of production mechanisms for heavy quarkonia~\cite{Celiberto:2022dyf,Celiberto:2022keu,Celiberto:2024omj}, rare~\cite{Celiberto:2025ogy}, and exotic hadrons~\cite{Celiberto:2023rzw,Celiberto:2024mab,Celiberto:2025ipt} within a unified collinear framework.

For the sake of illustration, Fig.~\ref{Fig:frags} display the $z$-dependence of the TQ4Q1.1 FFs directly obtained from the {\tt JETHAD} code~\cite{Celiberto:2020wpk,Celiberto:2022rfj}. 
They describe gluon (left panels) and charm (right panels) fragmentation into fully charmed tetraquarks with scalar ($T_{4c}(0^{++})$), axial-vector ($T_{4c}(1^{+-})$), and tensor ($T_{4c}(2^{++})$) quantum numbers.  
Each curve corresponds to a distinct factorization scale $\mu_F$, while the filled bands represent the propagated uncertainties associated with the non-perturbative long-distance matrix elements (LDMEs; for more details, see Sec.~II.C of Ref.~\cite{Celiberto:2025ziy}).
{Possible contributions from color-octet channels are neglected, as they are expected to be suppressed by at least two orders of magnitude relative to the singlet ones, due to both dynamical and hadronization constraints~\cite{Celiberto:2025ziy}. We also note that double-parton scattering, although relevant at low transverse momentum in multiquark final states~\cite{Maciula:2020wri}, is expected to be suppressed at moderate to large transverse momentum, where leading-power, single-parton fragmentation remains the dominant production mechanism.
}

To highlight the impact of perturbative evolution, the TQ4Q1.1 FFs are shown at representative factorization scales $\mu_F = 30$, $60$, and $90$~GeV, evolved from the input value $Q_0 = 5m_c$.  
At $Q_0$, the $[g \to T_{4c}(1^{+-})]$ function vanishes by construction, being generated radiatively through DGLAP evolution, whereas the scalar and tensor channels already feature nonzero gluon components.  
For uniformity, only evolved distributions are displayed.

The gluon-induced functions show broad support at small $z$. 
For scalar and tensor states, they decrease monotonically with increasing $z$ and lack any high-$z$ enhancement, while the axial-vector channel develops a distinct radiative hump in the intermediate region ($0.15 \lesssim z \lesssim 0.7$) that shifts slightly upward with increasing $\mu_F$.  
This feature, more evident in the charm sector, reflects the intrinsic shape dynamically generated by DGLAP evolution in the absence of a nonperturbative input.
The heavy-quark functions exhibit complementary patterns.  
In the scalar and tensor cases, they start from finite values at small $z$, drop rapidly, and develop a mild large-$z$ enhancement, consistent with hard-fragmentation behavior, typical of heavy-quarkonium production~\cite{Braaten:1993mp}.  
Conversely, the axial-vector channel peaks sharply at intermediate-to-large $z$ ($0.75 \lesssim z \lesssim 0.9$) and vanishes as $z \to 0$, revealing its strong preference for hard fragmentation and reduced soft-emission phase space.
In all cases, the axial-vector fragmentation probabilities are suppressed by over an order of magnitude relative to the scalar and tensor ones, in both gluon and quark channels.  
This hierarchy aligns with NRQCD expectations: the antisymmetric spin-color configuration of the $1^{+-}$ state yields smaller LDMEs and weaker overlap with leading gluon-production mechanisms, whereas the absence of orbital excitation enhances the scalar and tensor compatibility with collinear NRQCD selection rules.

\section{Results}
\label{Sec:Results}
In what follows, we will estimate the fragmentation contribution for the transverse momentum distribution of a $T_{4c}$ state produced in $pp$ collisions at $\sqrt{s} = 13$ TeV (LHC) and $\sqrt{s} = 100$ TeV (FCC, nominal). Our calculations will be performed considering different values of the rapidity $y$, assuming distinct PDF parameterizations and  taking into account of the GI and CI channels. Moreover,  we will present predictions for the 
three lowest Fock-state $T_{4c}$ configurations: scalar ($J^{PC}=0^{++}$), axial-vector ($J^{PC}=1^{+-}$), and tensor ($J^{PC}=2^{++}$). 
{The CMS Collaboration has recently reported~\cite{CMS:2025fpt} the first measurement of the quantum numbers of a family of fully charmed tetraquark candidates, with results being consistent with a $J^{PC} = 2^{++}$ assignment. This observation provides valuable support for the existence of tensor $T_{4c}$ states, but further studies are required to fully establish the quantum structure and possible existence of additional $T_{4c}$ states with different quantum numbers.}

\begin{figure}[t]
\includegraphics[scale=0.4]{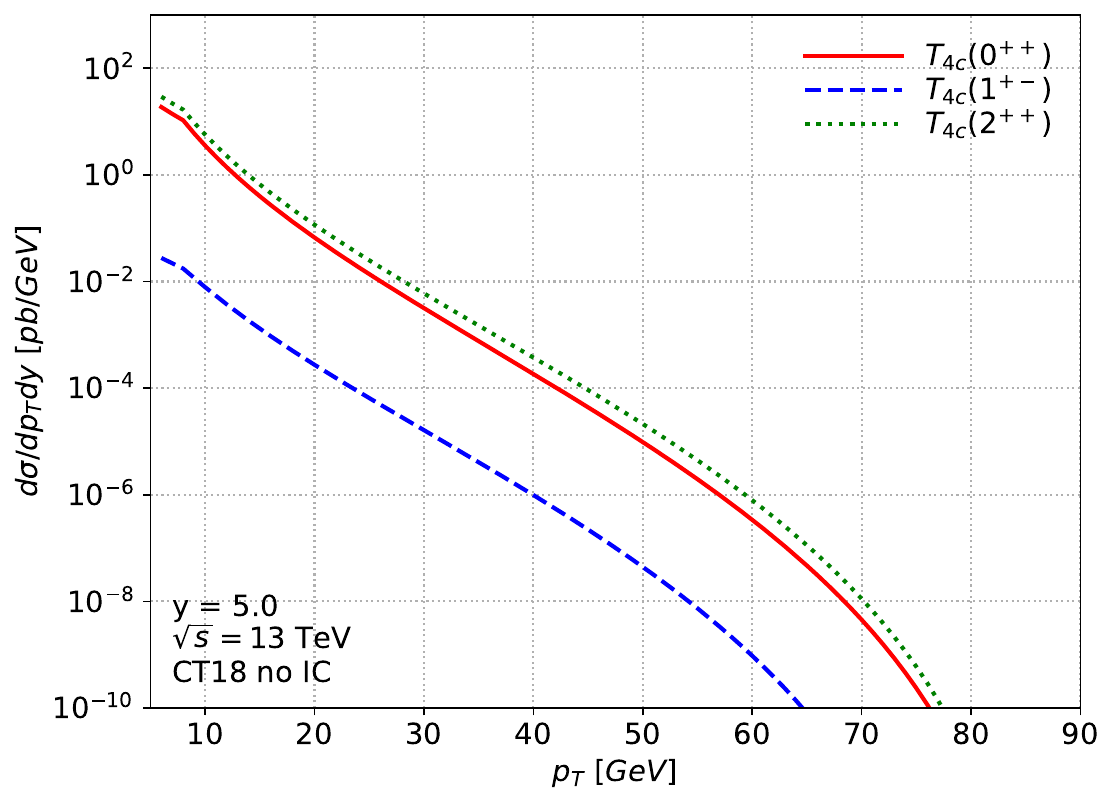}
\includegraphics[scale=0.4]{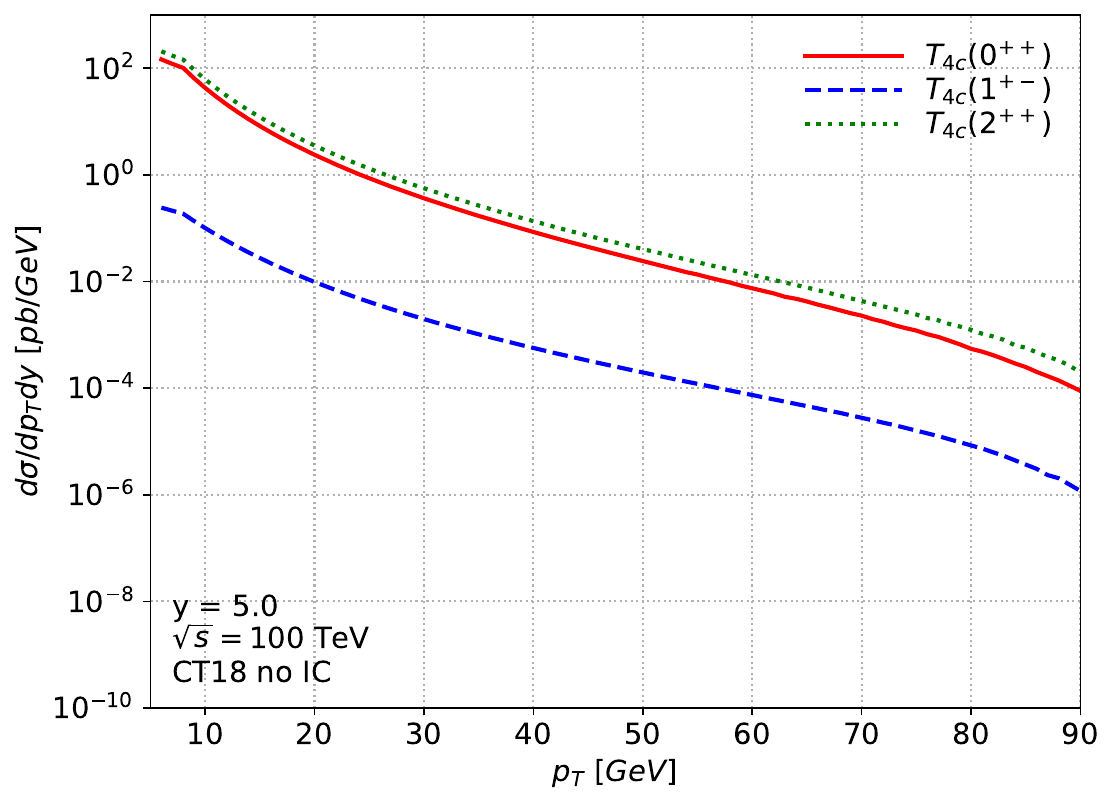}
\caption{Dependence of the transverse momentum distribution on the quantum numbers assumed for the $T_{4c}$ state. Results for the $T_{4c}$ production in $pp$ collisions at the LHC (left panel) and FCC (right panel) energies and fixed rapidity ($y = 5.0$).}
\label{Fig:ndependence}
\end{figure}

Initially, in Fig.~\ref{Fig:ndependence} we present the dependence of the transverse momentum distribution on the quantum numbers of the $T_{4c}$ state in $pp$ collisions at the LHC (left panel) and FCC (right panel) energies. The predictions have been derived considering the CT18 no IC parameterization and assuming $y = 5$, but the behaviors are observed for other choices. As expected, the cross - section increases with the center-of-mass energy. Moreover, our results indicate that the predictions for the $T_{4c}(2^{++})$ and $T_{4c}(0^{++})$ states are similar, while the production of a $T_{4c}(1^{+-})$ is suppressed by two orders of magnitude. Such suppression is directly associated with the smaller probability of a charm and gluon  fragmentation  into this state in comparison to scalar and tensor states, already observed in Fig.~\ref{Fig:frags}.
In what follows, we will focus on the $T_{4c}(2^{++})$ state.


\begin{figure}[t]
\includegraphics[scale=0.45]{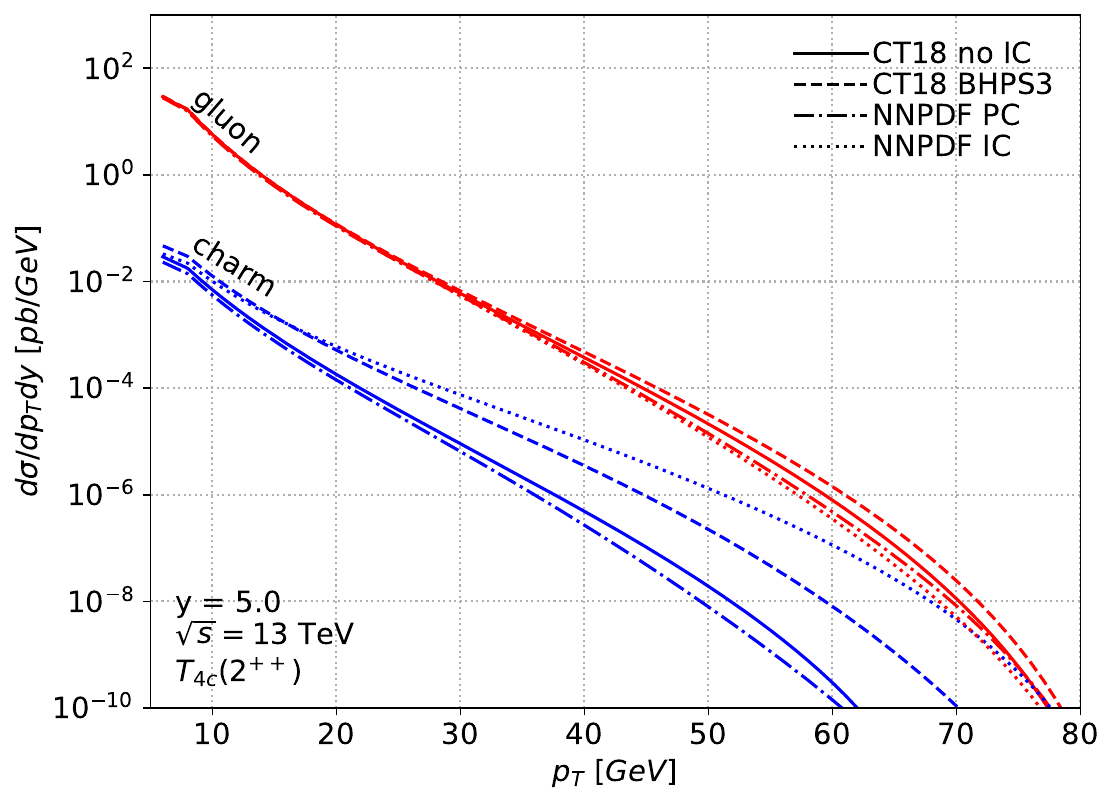}
\includegraphics[scale=0.45]{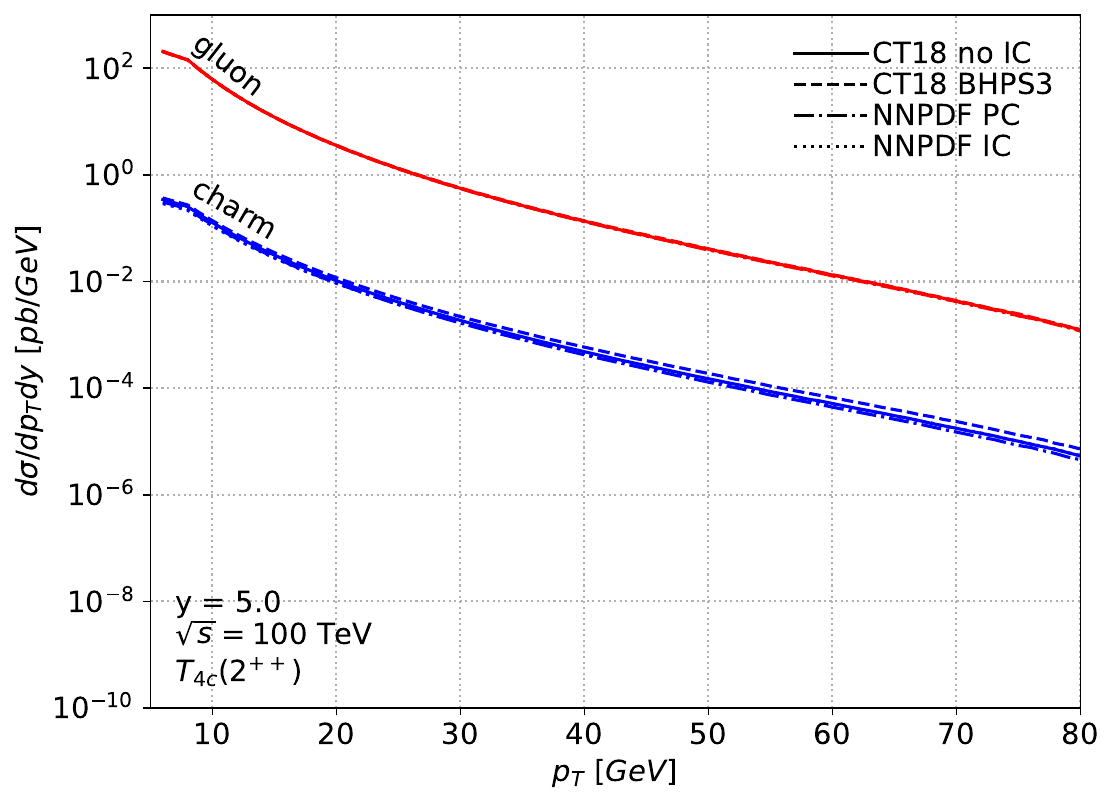}
\caption{Contributions of the GI and CI processes for the transverse momentum distribution associated with  the $T_{4c}(2^{++})$ production in $pp$ collisions at the LHC (left panel) and FCC (right panel) energies and fixed rapidity ($y = 5.0$). Results  derived considering the solution of the rcBK equation and distinct PDFs.}
\label{Fig:channels}
\end{figure}

In Fig.~\ref{Fig:channels} we present a comparison between the predictions associated with the contributions of the GI and CI processes for the transverse momentum distribution of a $T_{4c}(2^{++})$ state produced in $pp$ collisions at the LHC (left panel) and FCC (right panel) energies. The results have been derived considering four distinct PDF parameterizations, two of them taking into account of the intrinsic charm component. Initially, let us discuss the results for the LHC (left panel). As expected from Fig.~\ref{Fig:pdfs}, the magnitude of the CI channel is sensitive to the IC, with the NNPDF IC implying a larger enhancement. In particular, such parameterization implies that the charm and GI processes become of the same order at large transverse momentum $p_T$. At the FCC energy (right panel), the predictions are similar in the transverse momentum range considered, which is expected, since the impact of an IC component occurs at large-$x$, which is only reached at larger $p_T$ when the center-of-mass energy is increased.

\begin{figure}[t]
\includegraphics[scale=0.3]{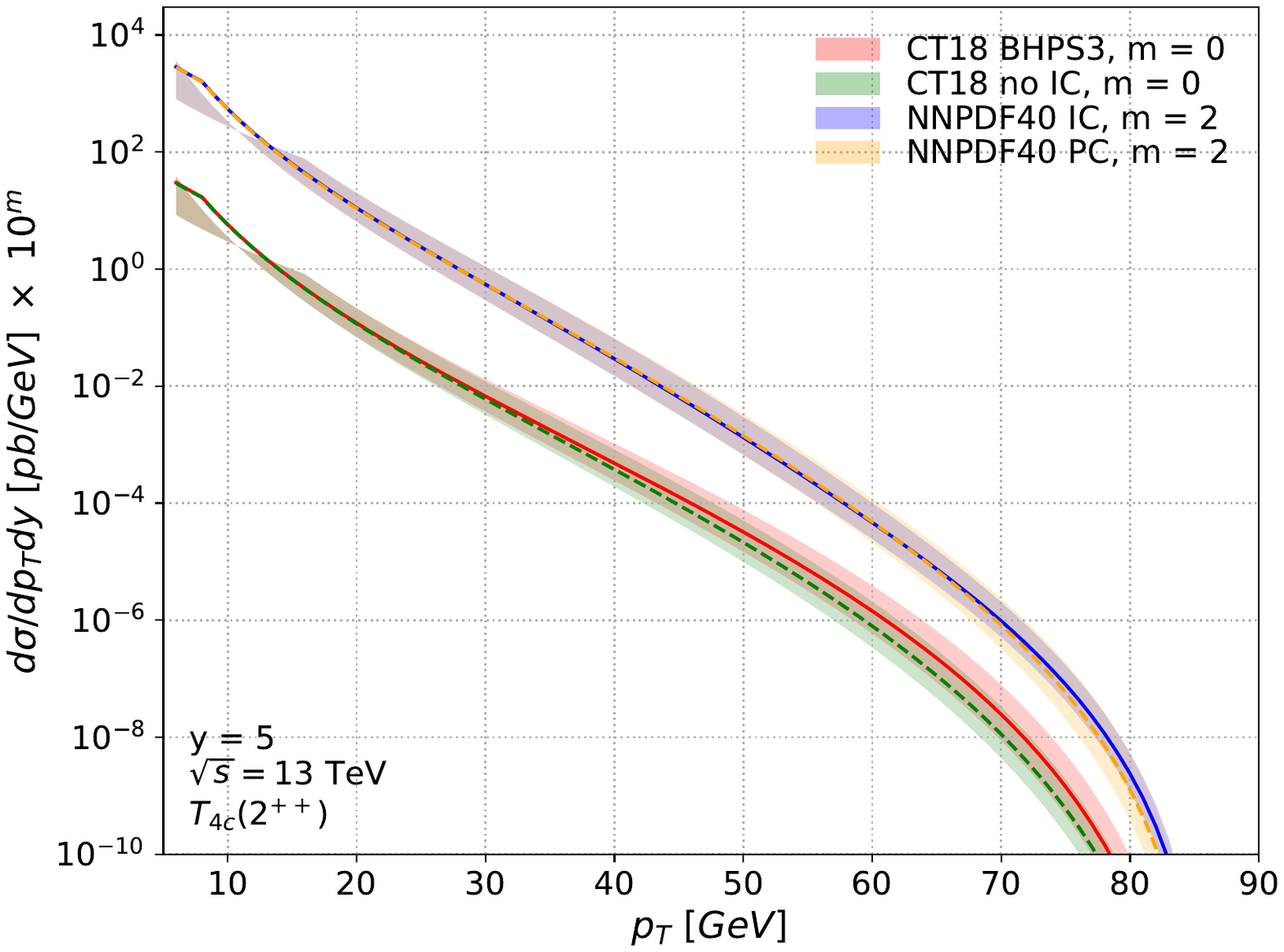}
\includegraphics[scale=0.3]{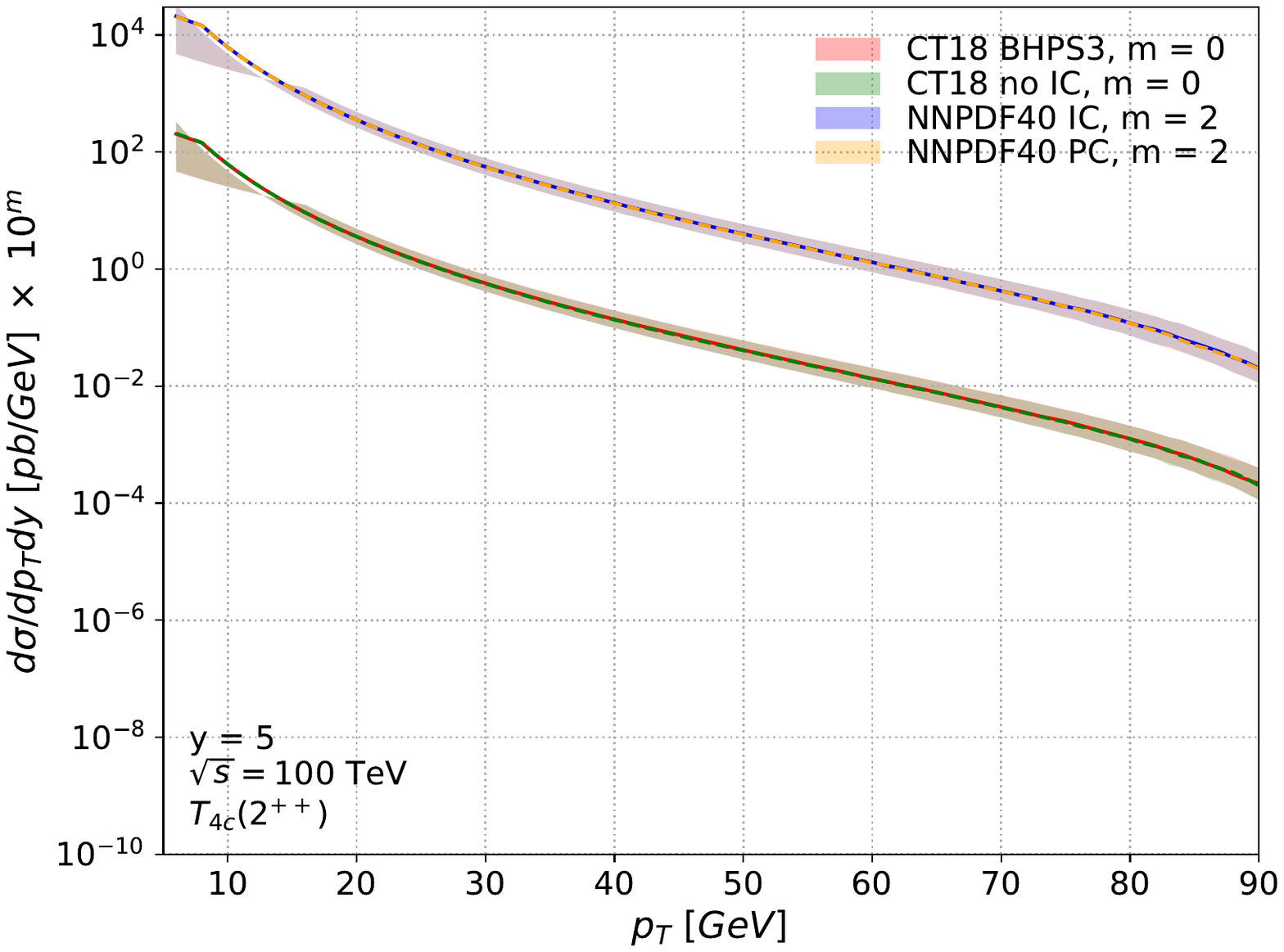}
\includegraphics[scale=0.3]{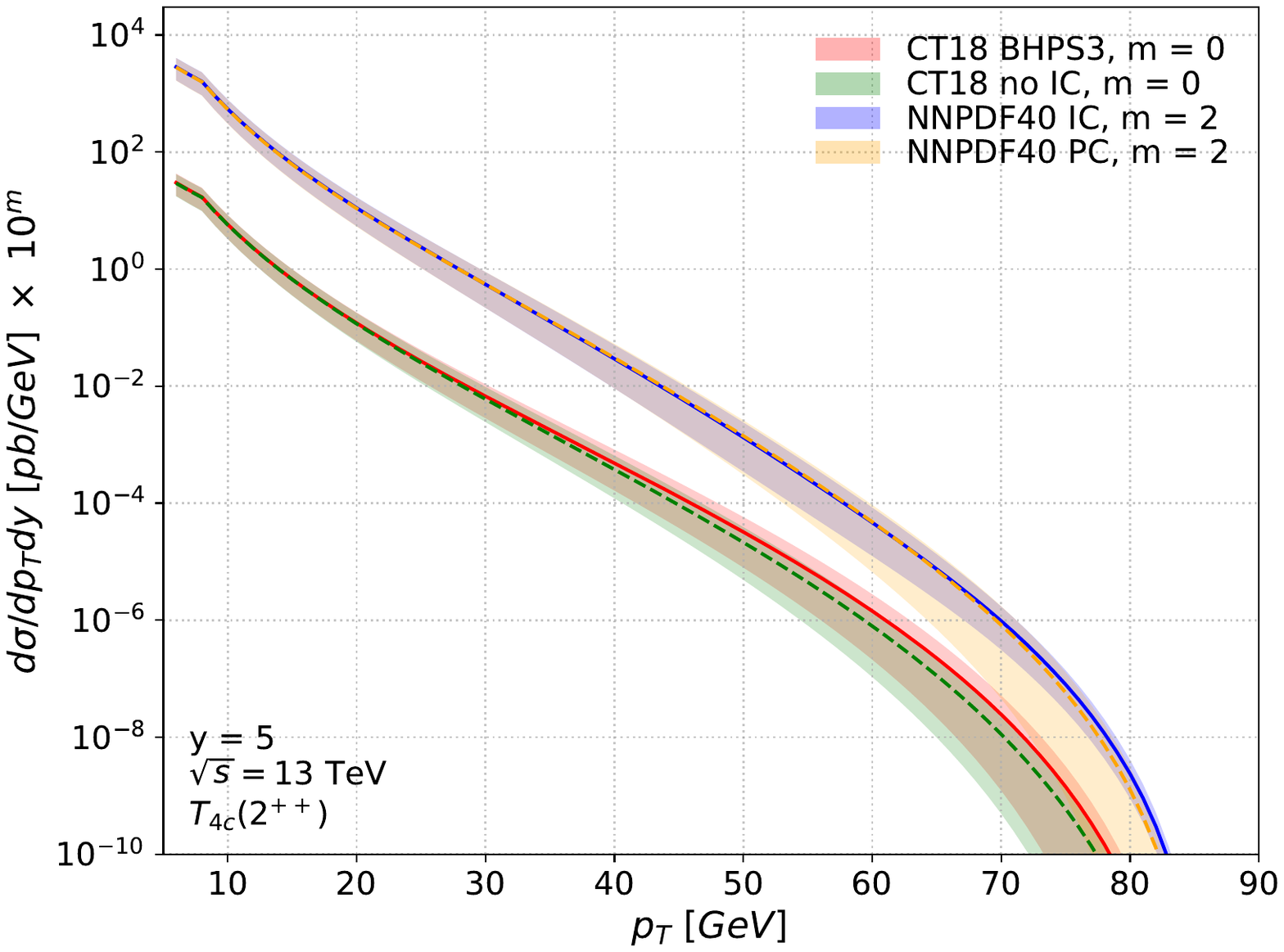}
\includegraphics[scale=0.3]{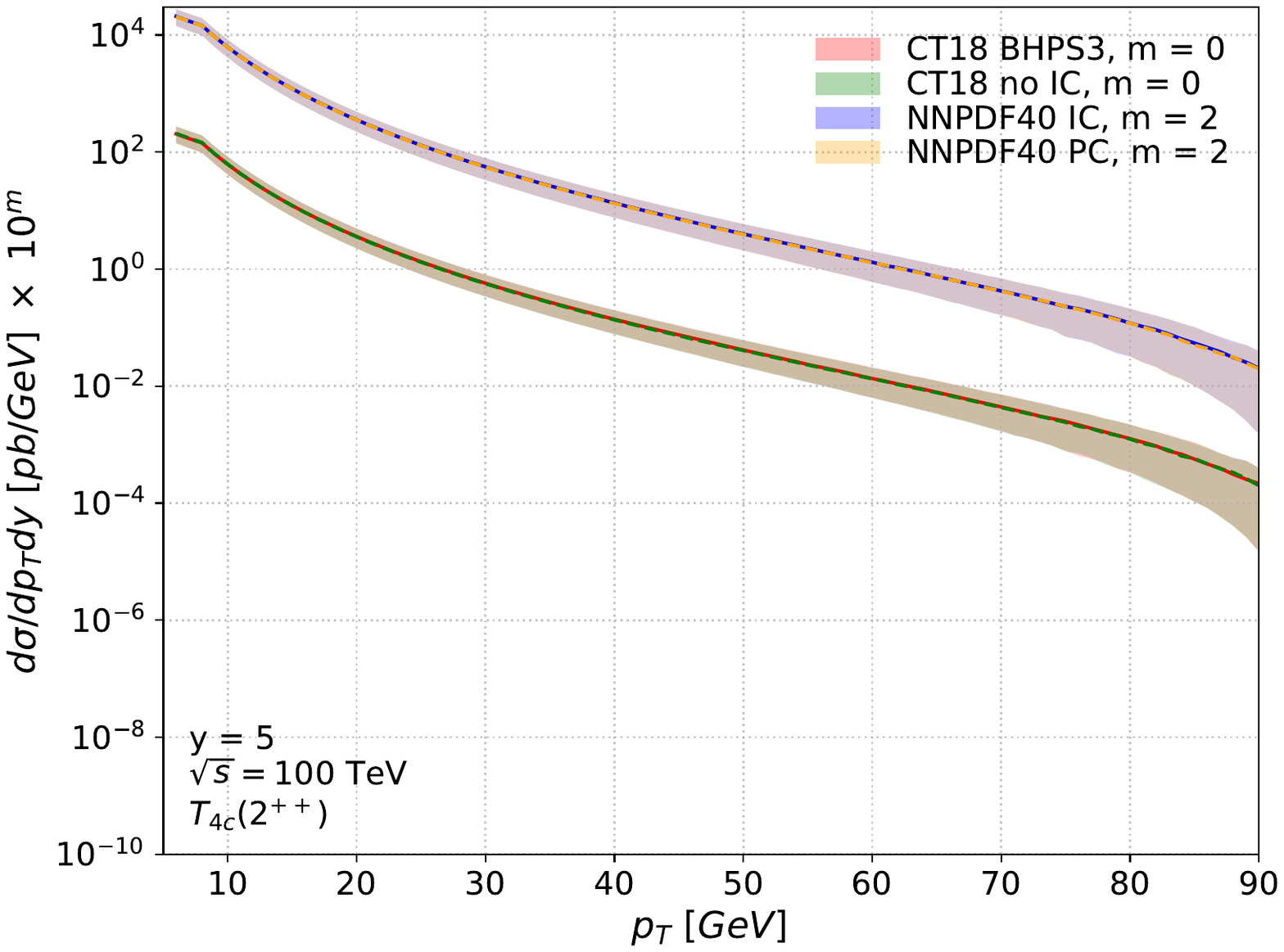}
\caption{ Dependence of our predictions for the transverse momentum distribution  associated with the $T_{4c}(2^{++})$ production in $pp$ collisions at the LHC (left panels) and FCC (right panels) energies on the factorization scale (upper panels) and non-perturbative long-distance matrix elements (lower panels). Results derived considering the solution of the rcBK equation and the CT18  and NNPDF  parameterizations.}
\label{Fig:uncertainties}
\end{figure}

{ In the previous figures for the transverse momentum distribution, we have assumed the central TQ4Q1.1 FFs and that the factorization and renormalization scales are equal and given by $\mu_R = \mu_F = \mu = M_{T_{4c}}$. Before  presenting  our predictions for distinct rapidities, it is important to evaluate the dependence of our results on these assumptions. In Fig. \ref{Fig:uncertainties} (upper panels) we illustrate the dependence on the factorization scale by varying $\mu$ in the range $\mu = M_{T_{4c}}/2 - 2 M_{T_{4c}}$. The results are presented for the LHC (left panel) and FCC (right panel) energies, different PDFs and a fixed rapidity ($y = 5$). We have that the uncertainty band increases at larger values of the transverse momentum and is smaller at the FCC energy in the $p_T$ range considered. In Fig. \ref{Fig:uncertainties} (lower panels), we present the dependence on the 
non-perturbative long-distance matrix elements (LDMEs), which determine the fragmentation functions. Our results indicate that this ingredient has a larger impact on our predictions. We have verified that  a similar conclusion is also obtained for other rapidities. As a consequence, in that follows, we will only present the uncertainty band associated to the LDMEs. 
}

\begin{figure}[t]
\includegraphics[scale=0.3]{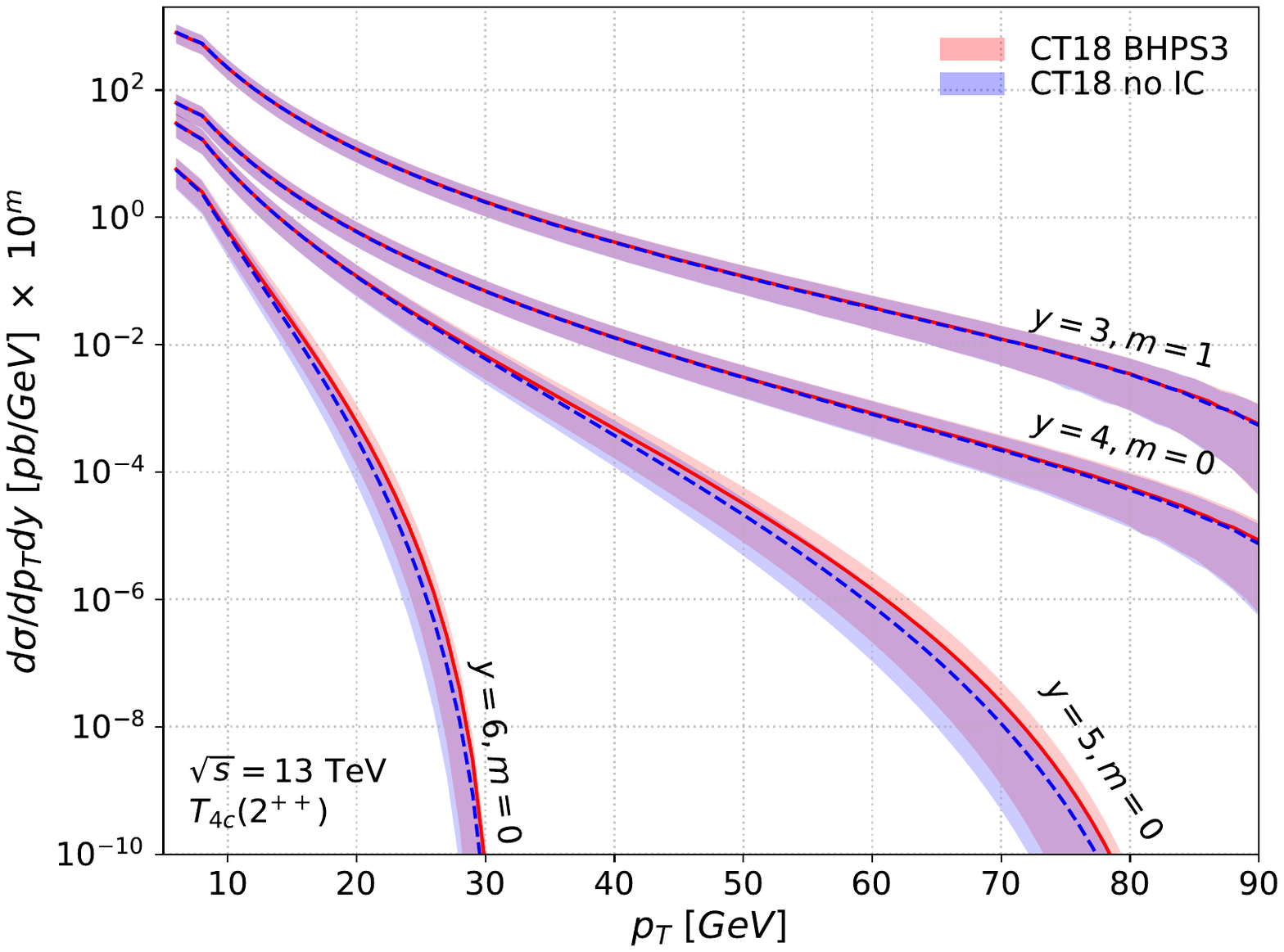}
\includegraphics[scale=0.3]{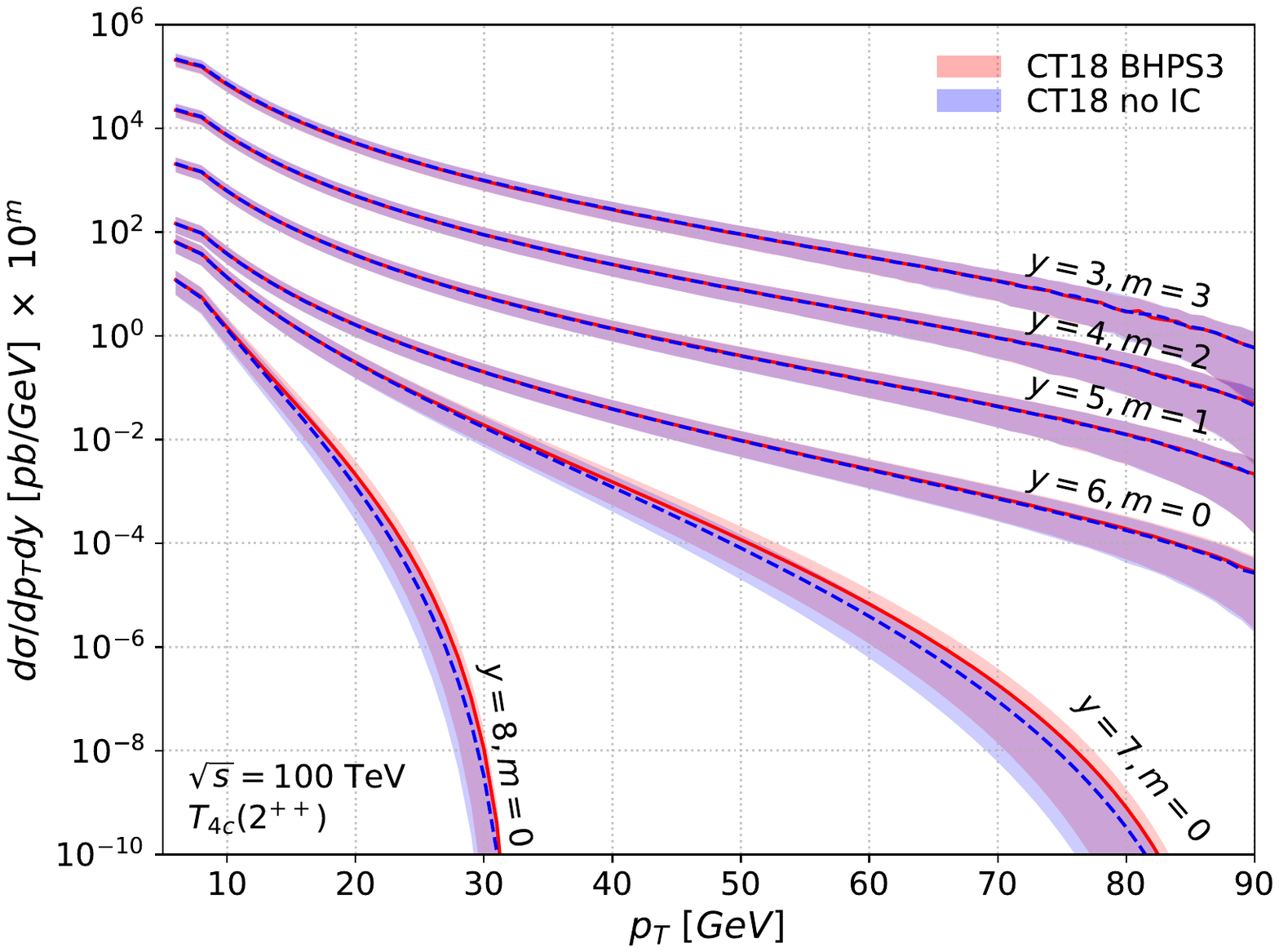}
\includegraphics[scale=0.3]{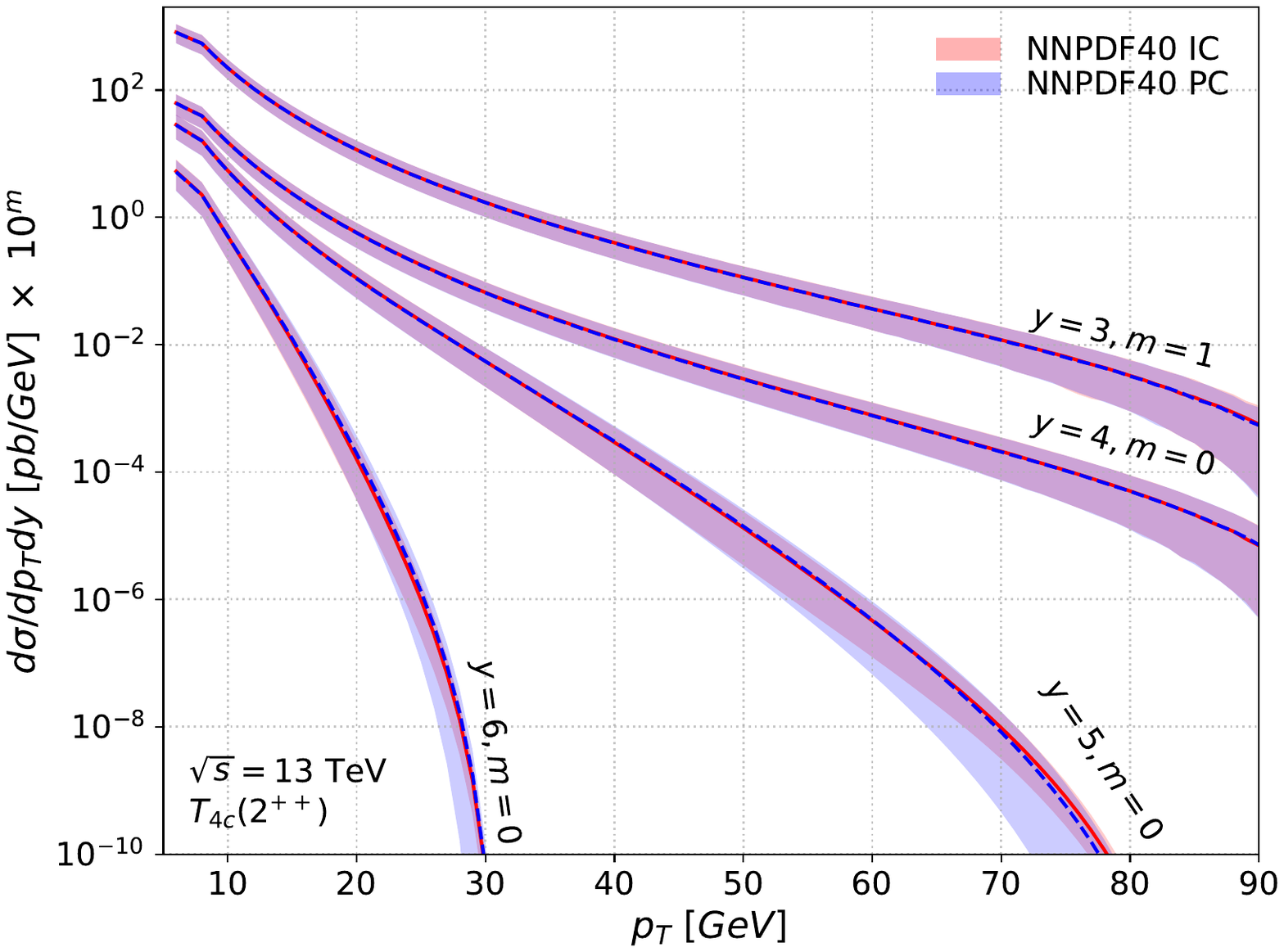}
\includegraphics[scale=0.3]{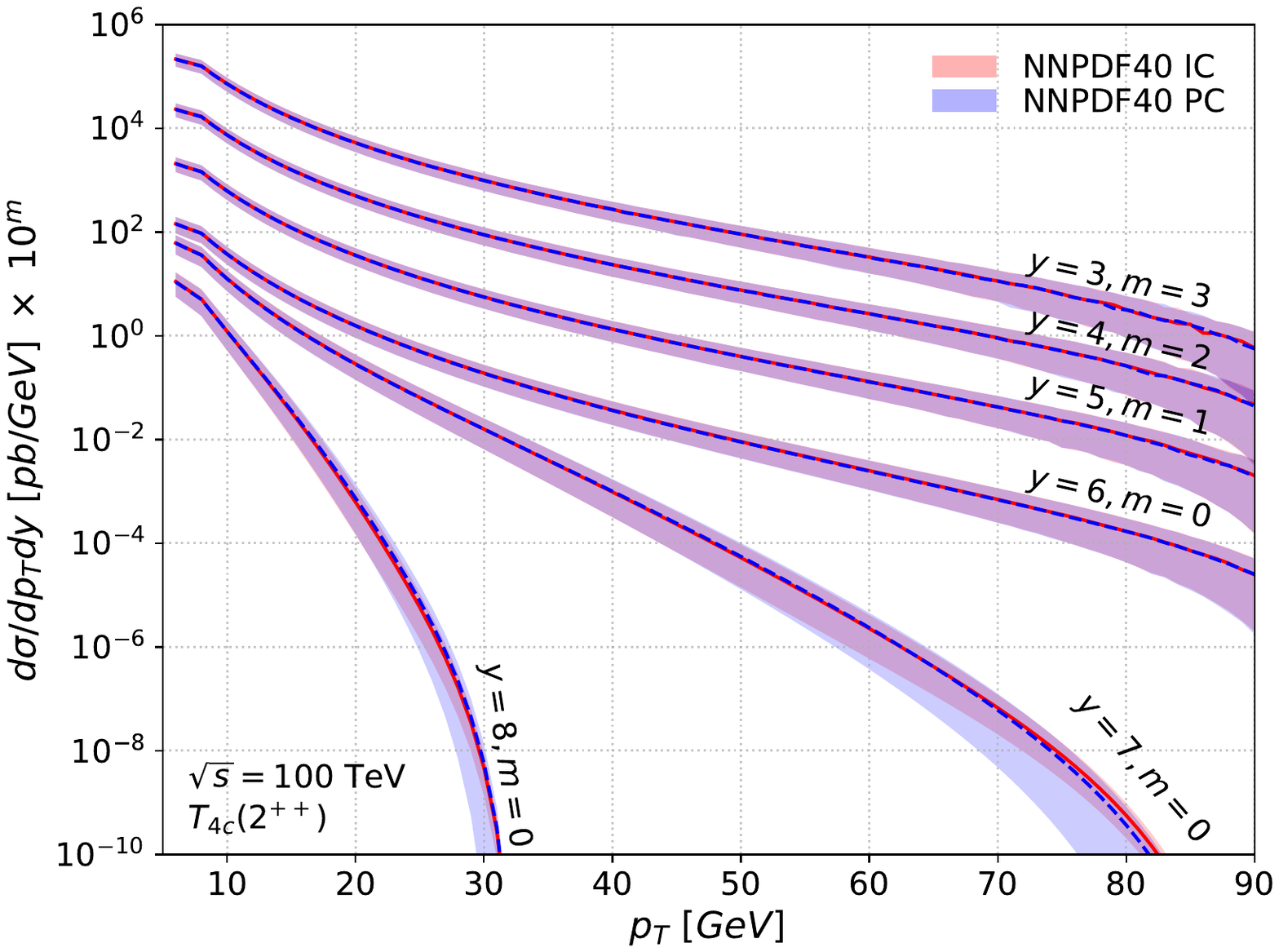}
\caption{Transverse momentum distribution for the $T_{4c}(2^{++})$ production in $pp$ collisions at the LHC (left panels) and FCC (right panels) energies and different rapidities, derived considering the solution of the rcBK equation and the CT18 (upper panels)  and NNPDF (lower panels)  parameterizations.}
\label{Fig:rapidities}
\end{figure}

The dependence of our predictions for the transverse momentum distributions    on the rapidity are presented in  Fig.~\ref{Fig:rapidities}. We consider 
 the $T_{4c}(2^{++})$ production in $pp$ collisions at the LHC (left panels) and FCC (right panels) energies, and the results have been derived by summing the GI and CI contributions and assuming different PDF parameterizations. The distribution increases with the energy and decreases with the rapidity. In particular, the behavior is strongly modified with the increasing of rapidity, due to the reduction of the available phase space for a fixed center-of-mass energy. Moreover, we have that the predictions associated with the distinct PDFs are similar, which is expected for this state due to the dominance of the GI channel.


\begin{figure}[t]
\includegraphics[scale=0.3]{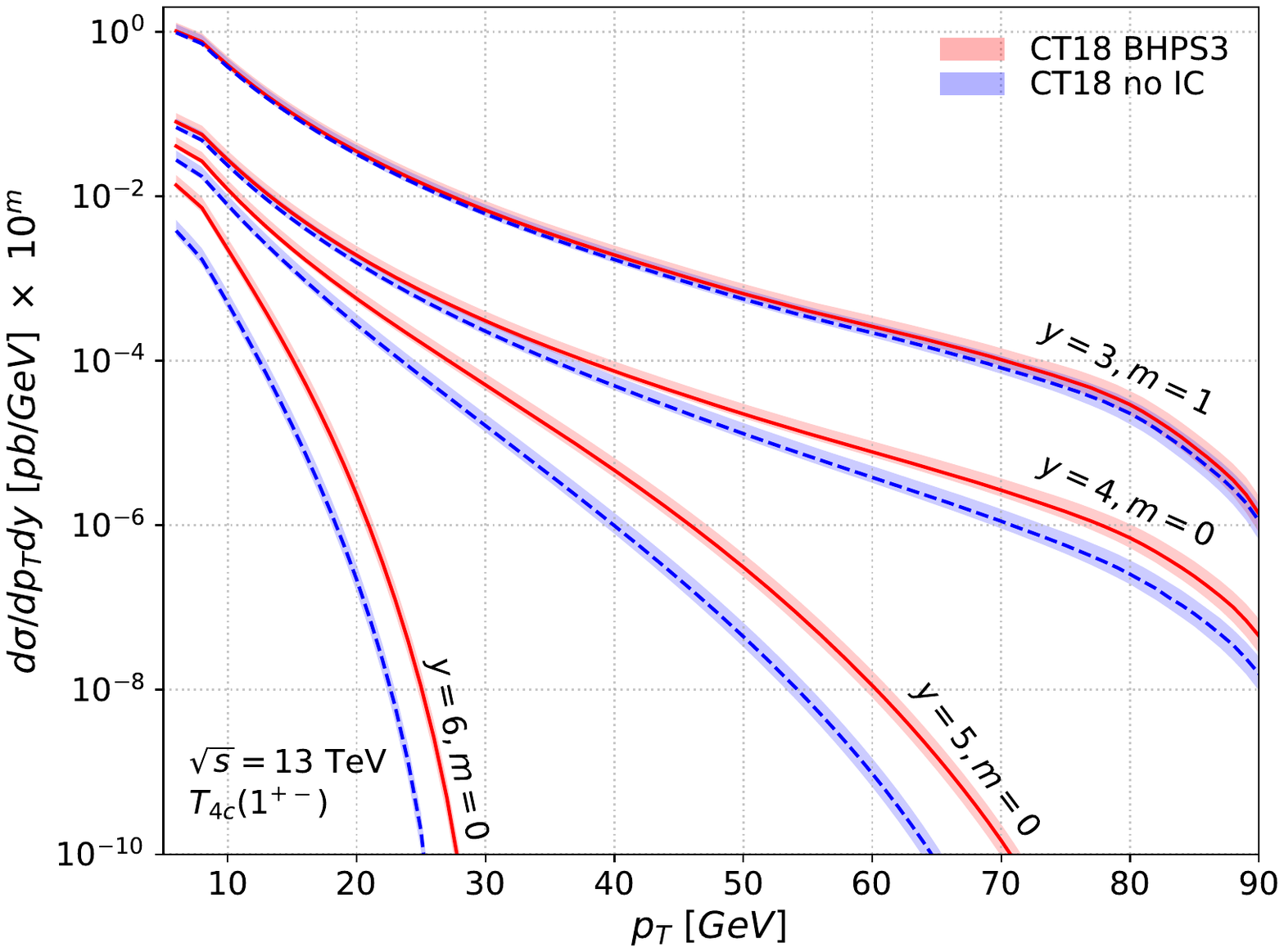}
\includegraphics[scale=0.3]{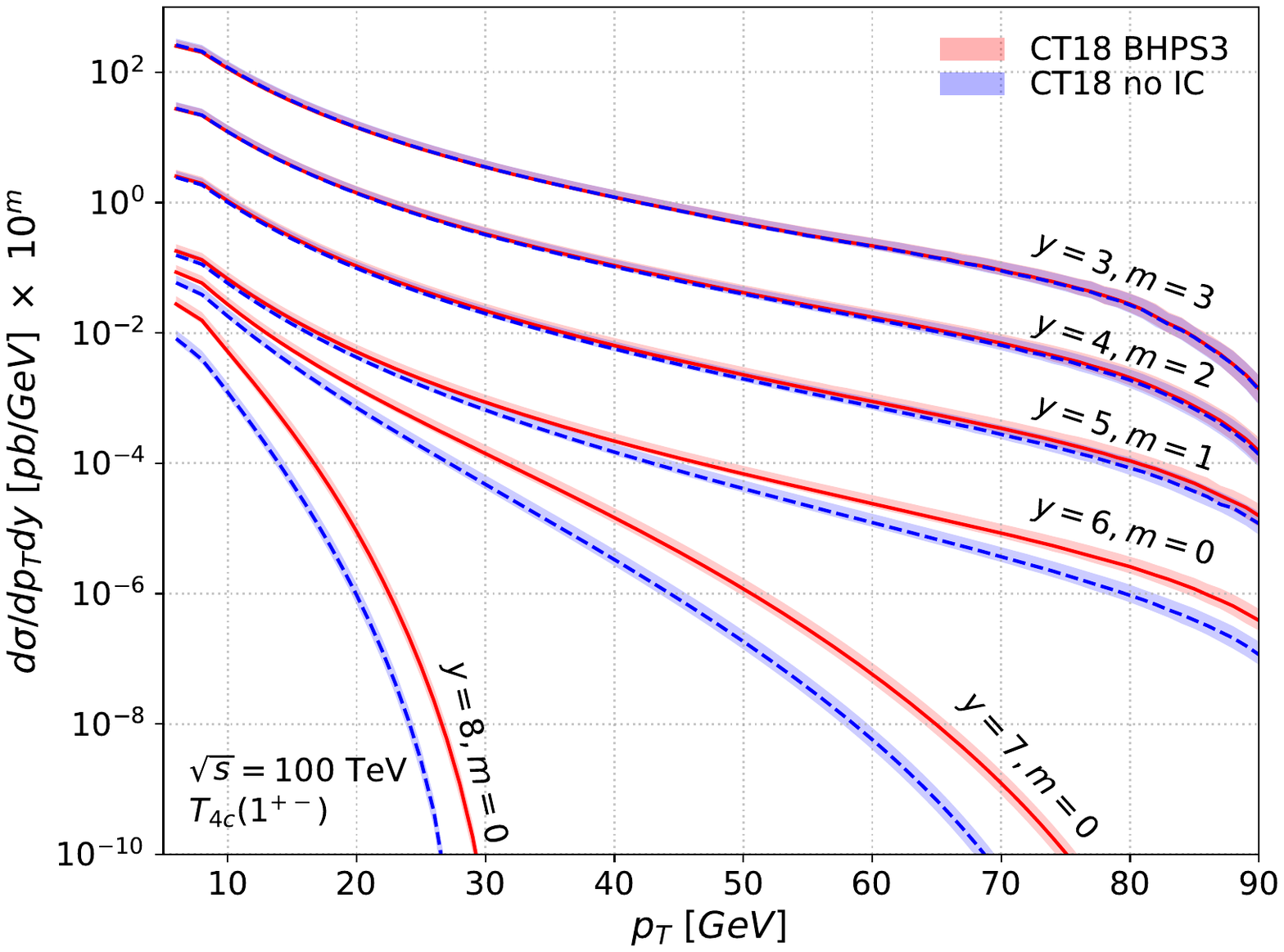}
\includegraphics[scale=0.3]{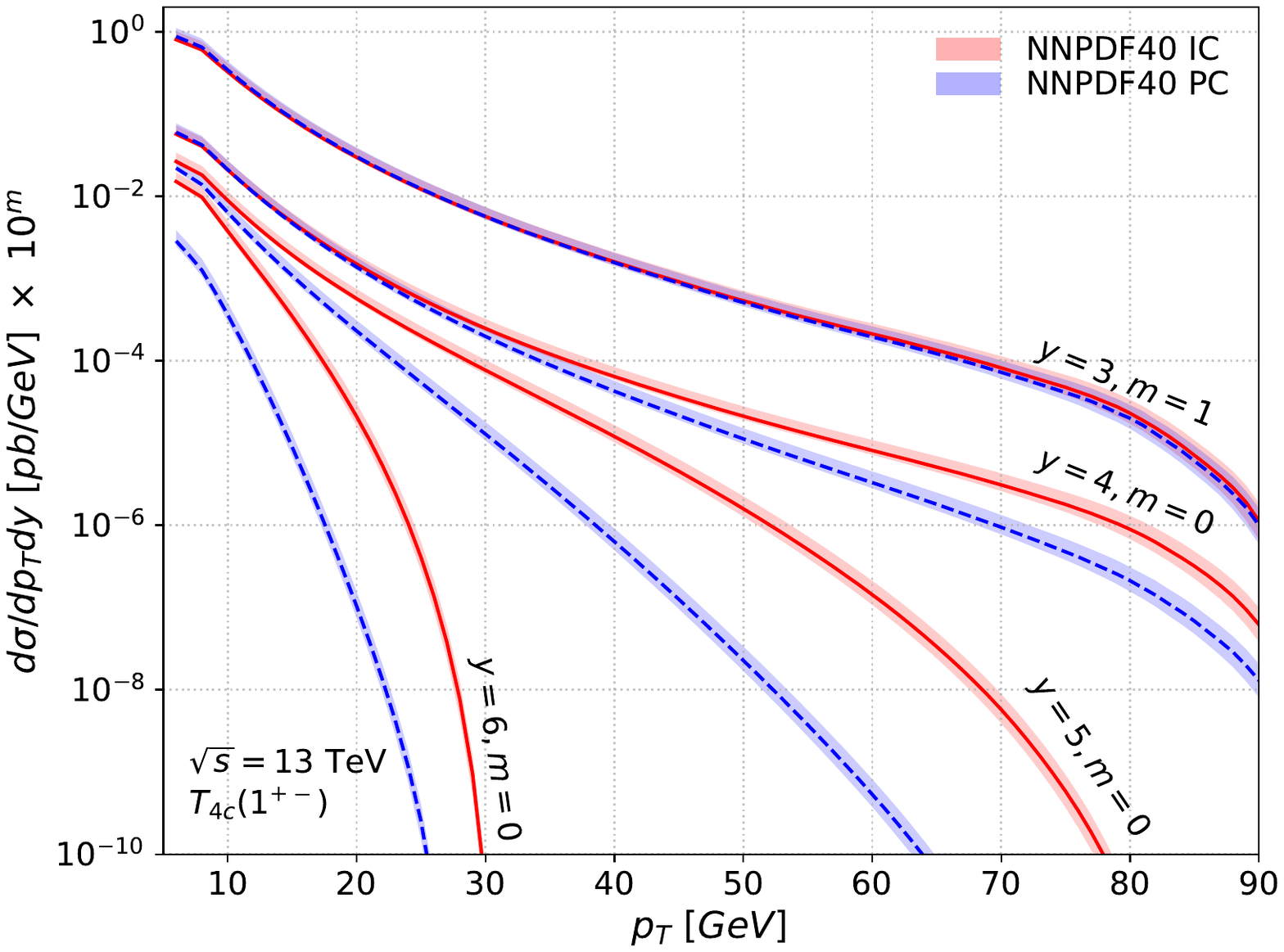}
\includegraphics[scale=0.3]{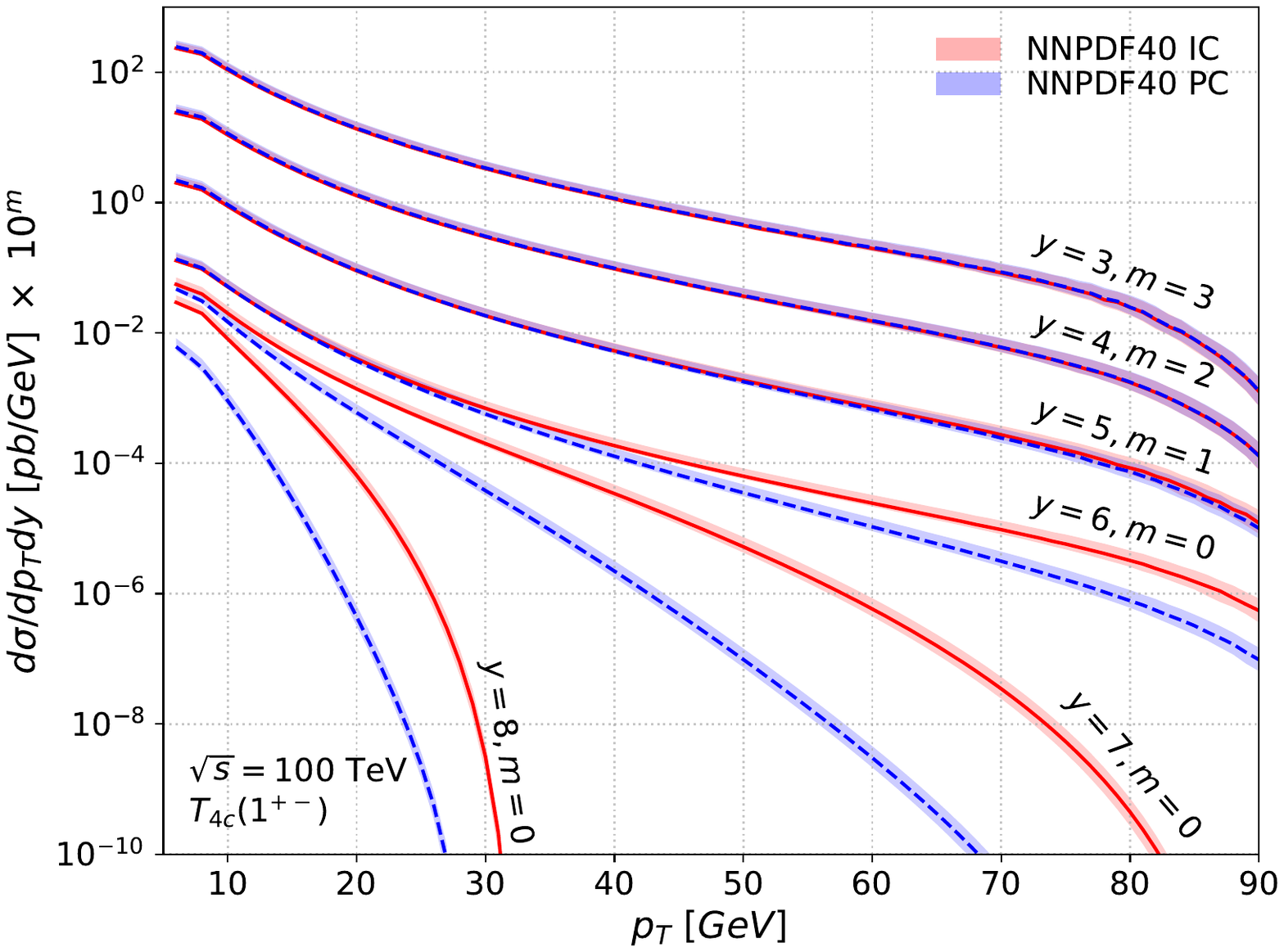}
\caption{Transverse momentum distribution for the $T_{4c}(1^{+-})$ production in $pp$ collisions at the LHC (left panel) and FCC (right panel) energies and different rapidities, derived considering the solution of the rcBK equation and the  CT18 (upper panels) and 
NNPDF  parameterizations. }
\label{Fig:axial}
\end{figure}

Let us now consider the production of a $T_{4c}(1^{+-})$ state. As verified in the previous section, the charm fragmentation is dominant for this state, differently from the scalar and tensor states, which makes the $T_{4c}(1^{+-})$ production more sensitive to an IC component in the proton wave function. In Fig.~\ref{Fig:axial} we present our predictions 
for the transverse momentum distribution of a $T_{4c}(1^{+-})$ state produced in $pp$ collisions at the LHC (left panel) and FCC (right panel) energies, derived  using the CT18 and NNPDF parameterizations. We have that the distribution  is strongly enhanced when an IC component is presented, especially at larger rapidities, with the magnitude of the enhancement being dependent on the PDF parameterization considered. 

\begin{center}
\begin{table}[t]
\begin{tabular}{|c|c|c|c|}
\hline 
\textbf{State} & PDF parametrization & $\sigma_{\rm LHC}$[pb] & $\sigma_{\rm FCC}$[pb] \\
\hline \hline

\multirow{4}{*}{$T_{4c}(0^{++})$}
& CT18 no IC     & 760.70 -- 1904.28 & 5257.51 -- 12106.06 \\
\cline{2-4}
& CT18 BHPS3     & 771.67 -- 1927.26 & 5212.69 -- 11976.89 \\
\cline{2-4}
& NNPDF40 PC     & 746.88 -- 1873.38 & 5225.36 -- 12037.64 \\
\cline{2-4}
& NNPDF40 IC     & 750.89 -- 1833.17 & 5260.52 -- 12133.35 \\
\hline \hline

\multirow{4}{*}{$T_{4c}(1^{+-})$}
& CT18 no IC     & 6.46 -- 9.14    & 43.21 -- 59.21   \\
\cline{2-4}
& CT18 BHPS3     & 7.33 -- 10.30    & 43.65 -- 59.72   \\
\cline{2-4}
& NNPDF40 PC     & 5.96 -- 8.46     & 41.56 -- 56.99   \\
\cline{2-4}
& NNPDF40 IC     & 6.20 -- 8.73    & 40.56 -- 55.69   \\
\hline \hline

\multirow{4}{*}{$T_{4c}(2^{++})$}
& CT18 no IC     & 1214.94 -- 2921.03 & 8105.39 -- 17857.29 \\
\cline{2-4}
& CT18 BHPS3     & 1231.31 -- 2954.36 & 7998.99 -- 17590.97 \\
\cline{2-4}
& NNPDF40 PC     & 1194.20 -- 2876.16 & 8078.51 -- 17791.60 \\
\cline{2-4}
& NNPDF40 IC     & 1200.73 -- 2891.00 & 8116.32 -- 17911.80 \\
\hline
\end{tabular}
\caption{Predictions for the production cross-sections of the different fully charmed tetraquark states for the LHC ($\sqrt{s} = 13$ TeV) and FCC ($\sqrt{s} = 100$ TeV). Results, in pb, derived assuming $p_T \ge 20$ GeV and $2.0 \le y \le 4.5$,  considering distinct PDF parameterizations { and taking into account of the uncertainty associated with the LDMEs}.}
\label{table:cross}
\end{table}
\end{center}

Finally, in Table~\ref{table:cross}, we present our predictions for the total cross - sections associated with the production of the different fully charmed tetraquark states in $pp$ collisions at the LHC and FCC energies through the fragmentation mechanism. The results, presented in pb, have been derived integrating over the transverse momentum $p_T$ of the $T_{4c}$ state in the range $p_T \ge 20$ GeV and over the forward rapidity range covered by the LHCb detector ($2.0 \le y \le 4.5$). Moreover, we present the predictions for the distinct PDF parameterizations used in our analyses { and taking into account of the uncertainty associated with the LDMEs}. As expected from the results for the transverse momentum distributions, the cross-section increases with the energy and is larger for the  $T_{4c}(2^{++})$ state. Moreover, for the LHC energy, the presence of an IC component implies a small enhancement of the cross-section in the rapidity range considered. In contrast, at the FCC, the predictions with and without an IC are very similar, which is expected, since the impact of an IC component for this energy only occurs at very large $p_T$, while the integrated cross-section is dominated by small values of transverse momentum. The results presented in Table~\ref{table:cross} indicate that the cross-section for the $T_{4c}(2^{++})$ production is of the order of 2 nb at the LHC, which implies $\gtrsim  10^9$ events if an integrated luminosity of 3000 fb$^{-1}$ is considered. As a consequence, a future experimental analysis of this final state is, in principle, feasible.




\section{Summary}
\label{Sec:conc}

Understanding exotic hadrons remains one of the central challenges in hadronic physics, with the production mechanisms of fully heavy tetraquark states being the subject of intense theoretical scrutiny.
Previous studies have analyzed the fragmentation-driven production of these states within the NRQCD formalism, demonstrating that this channel becomes dominant at large transverse momentum.
However, such analyses have predominantly focused on central rapidity configurations in $pp$ collisions.
In this work, we have extended the investigation to the forward rapidity regime, where additional effects are expected to shape the production rates.
In particular, forward kinematics probes asymmetric Bjorken-$x$ configurations ($x_1 \gg x_2$), thereby enhancing sensitivity to non-linear QCD dynamics at small $x$ and to the large-$x$ structure of the projectile.
To model this dilute–dense scattering environment, we employed hybrid factorization for both gluon-initiated and charm-initiated channels, using the CGC formalism to describe the parton–target interaction.
We further incorporated different models of the intrinsic component in the proton wave function, and provided predictions for the production of $T_{4c}$ states with different quantum numbers.
Our phenomenological analysis included transverse momentum distributions at various rapidities and collider energies, ranging from the 13~TeV~LHC to the nominal 100~TeV~FCC.
We found that the largest cross-sections are associated with tensor states, predominantly produced via the gluon-initiated mechanism.
In contrast, axial-vector states are dominantly produced via charm-initiated fragmentation, making them particularly sensitive to the large-$x$ charm content of the projectile.
As such, the $T_{4c}(1^{+-})$ fragmentation production~\cite{Celiberto:2025dfe} emerges as a valuable probe of the intrinsic charm hypothesis, especially in forward kinematics where charm-initiated contributions gain greater relevance.

\section*{Acknowledgements}
V.P.G. was partially supported by CNPq,  FAPERGS and INCT-FNA (Process No. 408419/2024-5). Y.N.L. was partially financed by the São Paulo Research Foundation (FAPESP), Brazil (Process Number 2024/17836-9). A.V.G. is grateful to Universidade do Estado de Santa Catarina for its hospitality and financial support 
and is partially supported by CNPq through the INCT-FNA grant 408419/2024-5 and the Universal Grant 405458/2025-8. The authors acknowledge the National Laboratory for Scientific Computing (LNCC/MCTI, Brazil), through the
ambassador program (UFGD), subprojects FCNAE and SADFT for providing HPC resources of the SDumont super-
computer.
F.G.C. was supported by the Atracción de Talento Grant No. 2022-T1/TIC-24176 from the Comunidad Autónoma de Madrid, Spain.


\end{document}